\documentclass[preprint,12pt]{elsarticle}
\usepackage{amsmath}
\usepackage{amssymb}
\usepackage{graphicx}
\usepackage{subfigure}

\journal{Journal of Computational Physics}

\begin{document}

\begin{frontmatter}
  
  \title{The effective conductivity of arrays of squares: large random
    unit cells and extreme contrast ratios}

\author{Johan Helsing\fnref{label2}}

\fntext[label2]{Tel.:+46 46 2223372}

\ead{helsing@maths.lth.se}
\ead[url]{http://www.maths.lth.se/na/staff/helsing/}

\address{Numerical Analysis, Centre for Mathematical Sciences,\\ Lund
  University, Box 118, SE-221 00 LUND, Sweden}

\begin{abstract}
  An integral equation based scheme is presented for the fast and
  accurate computation of effective conductivities of two-component
  checkerboard-like composites with complicated unit cells at very
  high contrast ratios. The scheme extends recent work on
  multi-component checkerboards at medium contrast ratios. General
  improvement include the simplification of a long-range
  preconditioner, the use of a banded solver, and a more efficient
  placement of quadrature points. This, together with a reduction in
  the number of unknowns, allows for a substantial increase in
  achievable accuracy as well as in tractable system size. Results,
  accurate to at least nine digits, are obtained for random
  checkerboards with over a million squares in the unit cell at
  contrast ratio $10^6$. Furthermore, the scheme is flexible enough to
  handle complex valued conductivities and, using a homotopy method,
  purely negative contrast ratios. Examples of the accurate
  computation of resonant spectra are given.
\end{abstract}


\end{frontmatter}

\section{Introduction}

This paper is devoted to solving the electrostatic equation for
periodic composites with unit cells made of squares of conductivity
$\sigma_2$ that are either mixed with other squares of conductivity
$\sigma_1$, to form a checkerboard structure, or simply embedded in a
background material of conductivity $\sigma_1$. There are $N_{\rm sq}$
squares in the unit cell and the area fraction of squares with
conductivity $\sigma_2$ is denoted $p$. The goal is to compute the
effective conductivity $\sigma_{\ast}$ rapidly, with high accuracy,
and for almost any combination of $\sigma_1$, $\sigma_2$, and $p$ for
which the electrostatic equation has a solution.

\subsection{Motivation and challenges}
\label{sec:motif}

There are several applications that motivate our study. The
homogenization of checkerboards with random unit cells at high real
valued contrast ratios $\sigma_2/\sigma_1$ (or $\sigma_1/\sigma_2$) is
a classic problem in materials science. It is of interest to study how
$\sigma_{\ast}$ depends on $p$ and in particular what happens when one
type of squares forms a connected path throughout the composite
(percolation). The contrast ratio can be considerable in materials of
technological importance. A ratio of $10^7$ is not
unusual~\cite{Mart03}. Very large $N_{\rm sq}$ are then needed to
reach convergence to statistical limits. See Chapters 10.10 and 10.11
of~\cite{Milt02} for a review of this field. In the metamaterial
community there is a strong interest in a related issue, namely how to
compute resonant spectra of effective dielectric permittivity
functions (spectra of plasmonic excitations) for composites made of
polygonal metamaterial inclusions embedded in a dielectric background
material~\cite{Fred03,Perr10}. The electrostatic equation is the same
for conducting and for dielectric materials. Only the notation
differs, see Table on p.~19 of~\cite{Milt02}. For simplicity, we will
talk about conductivity in this context, too. Of particular interest
is the behavior of $\sigma_{\ast}$ close to values of
$\sigma_2/\sigma_1$ where the electrostatic equation does not have a
solution or only has a solution as a limit in the complex
$\sigma_2/\sigma_1$-plane.

The computational tasks just discussed offer extreme challenges.
Non-smooth interfaces tend to make solutions singular and hard to
resolve. The electric fields close to certain corner vertices may just
barely be square integrable. As $N_{\rm sq}$ grows, the interaction
between distantly separated parts in the computational domain may
cause problems which cannot be resolved by discretization and local
techniques alone. Being in the vicinity of parameter combinations
where the electrostatic equation ceases to have a solution is often
hard. All these difficulties add up and may manifest themselves as
artificial ill-conditioning, slow convergence with mesh refinement,
critical slowing down in iterative solvers, and severe loss of
precision. Several methods have been suggested to alleviate these
problems including variants of the finite element
method~\cite{Berg01,Chen09}, network models~\cite{Mart03,Hamo09},
renormalization schemes~\cite{Kari10}, mode-matching
methods~\cite{Perr10}, and Brownian motion simulation~\cite{Kim03}.
See also Section~3 of~\cite{Nuka07} for state-of-the-art algorithms to
combat critical slowing down in network models and~\cite{Gill11} for a
discussion of future directions in the research field at large.

\subsection{Our scheme}

Let $\Gamma$ denote the boundary (the interfaces) of a composite. We
shall reformulate the electrostatic equation as a Fredholm second kind
integral equation
\begin{equation}
\left(I+K\right)\mu(z)=g(z)\,, \quad z\in\Gamma\,,
\label{eq:gen0}
\end{equation}
where $I$ is the identity, $K$ is an integral operator which is
compact on smooth $\Gamma$, $\mu(z)$ is an unknown layer density,
and $g(z)$ is a right hand side.

Solvers for large-scale boundary values problems on smooth domains
often rely on integral equation reformulations of the
form~(\ref{eq:gen0}). The last few years have seen increased activity
in the development of efficient solvers using~(\ref{eq:gen0}) also
when $\Gamma$ is non-smooth. The scheme of the present paper
originates from work on non-smooth inclusion problems in
free-space~\cite{Hels08}. The ideas in~\cite{Hels08} were later
improved and extended to encompass the biharmonic
equation~\cite{Hels09a}, mixed boundary conditions~\cite{Hels09b},
singular integral equations with non-zero indices~\cite{Hels11a}, and
boundaries with quadruple-junctions~\cite{Hels11b}. The present paper
is a direct sequel to~\cite{Hels11b}. As in~\cite{Hels11b}, we apply a
combination of short- and long-range preconditioners
to~(\ref{eq:gen0}). Major new features include:
\begin{itemize}
\item A better strategy for choosing quadrature nodes which makes the
  error in $\sigma_{\ast}$ grow linearly with contrast ratio.
  In~\cite{Hels11b} the growth is superlinear.
\item An improved long-range preconditioner which makes the
  computational cost grow almost linearly with $N_{\rm sq}$.
  In~\cite{Hels11b} the growth is cubic.
\item A homotopy-type method which allows for computing
  $\sigma_{\ast}$ at points in the complex $\sigma_2/\sigma_1$-plane
  where the solution to the electrostatic equation only exists as a
  non-unique limit.
\end{itemize}
In addition there are several minor improvements.

\subsection{Relation to the Bremer--Rokhlin scheme}

Other recent work on the efficient solution of~(\ref{eq:gen0}) in the
presence of non-smooth boundaries includes~\cite{Brun09}, which
exploits cancellation of singularities, and a comprehensive mechanism
currently being developed by a group around Bremer and
Rokhlin~\cite{Brem10b,Brem10a,Brem10c}. Let us discuss the relation of
the Bremer--Rokhlin scheme to our scheme.

Both schemes take as a starting point the observation that an accurate
and economical discretization of~(\ref{eq:gen0}) can only be effected
by restricting the operator on the left hand side to a
finite-dimensional subspace, determined by the right hand side $g(z)$.
High resolution in combination with compression is used as a means to
achieve this. The result is a kind of precomputed purpose-made
composite quadrature, suitable for Nystr{\"o}m discretization.

The Bremer--Rokhlin scheme employs an elaborate machinery to construct
families of `universal quadratures'. Each universal quadrature is
appropriate for the discretization of a given integral equation over
an entire class of boundary segments with complicated geometry. When
the integral equation depends on material parameters, in addition to
geometry, more universal quadratures are needed. The approach has the
advantage that the precomputation is done once and for all and can be
stored on disk. When solving a particular problem involving many
boundary singularities of similar shapes, only a few universal
quadratures need to be activated. Our scheme precomputes
`quadrature-weighted inverses' afresh around every boundary
singularity. This offers greater flexibility when applying the scheme
to new situations and opens up for a parallel implementation, but
requires more RAM storage.

Another difference between the two schemes is the way in which the
process of resolution and compression is carried out. The
Bremer--Rokhlin compression is done via a series of solutions of large
linear systems followed by rank-revealing $QR$ decompositions. Our
scheme deals with resolution and compression in tandem, using a fast
and stable recursion. No large linear systems are ever set up. This is
an advantage for boundary segments where extremely high resolution is
needed.

The schemes also differ in the assumptions made on $g(z)$. The
Bremer--Rokhlin scheme assumes that $g(z)$ is a restriction to
$\Gamma$ of a function that satisfies the underlying partial
differential equation in a neighborhood of each point on $\Gamma$.
This assumption applies, for example, to certain important acoustic
scattering problems. Our scheme only assumes that $g(z)$ is piecewise
smooth. This is an advantage when the computational domain models
granular materials or materials containing branching cracks.

An open question is how easily the two schemes generalize to three
dimensions. Perhaps one can combine their best features?

\subsection{Organization of the paper}

The paper is divided into eight sections. Section~\ref{sec:inteq}
introduces unit cells and integral equations that will be used in all
examples. Section~\ref{sec:disc} is on discretization. The leading
ideas in our compression scheme are summarized in
Sections~\ref{sec:short} and~\ref{sec:comp}. Section~\ref{sec:recur}
is on implementation. This material is essential for the understanding
of how limits are taken in the complex $\sigma_2/\sigma_1$-plane and
how the compression of inverses of giant matrices corresponding to
intensely resolved integral operators can be executed in sub-linear
time. Section~\ref{sec:long} presents improvements to the long-range
preconditioner proposed in~\cite{Hels11b}. The paper ends in
Section~\ref{sec:numex} with some truly large-scale and accurate
numerical examples for random checkerboards along with the computation
of resonant spectra of two metamaterial composites. The reader
interested in more examples is referred to a forthcoming
paper~\cite{Hels11c}.

\section{Integral equations for the electrostatic problem}
\label{sec:inteq}

We shall solve the electrostatic partial differential equation on
three types of doubly-periodic domains in a plane $D$: square arrays
of squares, staggered arrays of squares, and two-component random
checkerboards. An average electric field $e=(e_x,e_y)$ of unit
strength is applied to $D$ and we seek the potential $U(r)$ for the
computation of $\sigma_{\ast}$ in direction $e$
\begin{equation}
\sigma_{\ast}=
\int_{D_0}\left(\sigma(r)\nabla U(r)\cdot e\right)\,{\rm d}V_r\,,
\label{eq:eff0}
\end{equation}
where $\sigma(r)$ is the local conductivity, ${\rm d}V_r$ is an
infinitesimal area element, and the unit cell $D_0$ is
$[-1/2,1/2)\times[-1/2,1/2)$. We make no distinction between points or
vectors in a real plane $\mathbb{R}^2$ and points in a complex plane
$\mathbb{C}$. From now on, all points will be denoted $z$ or $\tau$.

The interfaces $\Gamma$ in $D$ are given orientation. The restriction
of $\Gamma$ to $D_0$ is denoted $\Gamma_0$ and the outward unit normal
of $\Gamma$ at $z$ is $n_z=n(z)$. Corner vertices are denoted
$\gamma_k$. Obviously, $\sigma(z)$ may jump as $\Gamma$ is crossed.
Let $\sigma_+(z)$ denote the conductivity on the positive side of
$\Gamma$ at $z$, let $\sigma_-(z)$ denote the conductivity on the
negative side, and introduce as in~\cite{Hels11b}
\begin{align}
a(z)&=\sigma_+(z)-\sigma_-(z)\,,\quad z\in\Gamma\,,\\
b(z)&=\sigma_+(z)+\sigma_-(z)\,,\quad z\in\Gamma\,,\\
c(z)&=\sigma_+(z)\sigma_-(z)\,,\quad z\in\Gamma\,,\\
\lambda(z)&=a(z)/b(z)\,,\quad z\in\Gamma\,.
\label{eq:lmb}
\end{align}

Our domains exhibit similarities, but they also differ in important
respects. Different integral equation reformulations will be used for
efficiency.

\begin{figure}[t]
\begin{center}
  \includegraphics[width=32mm]{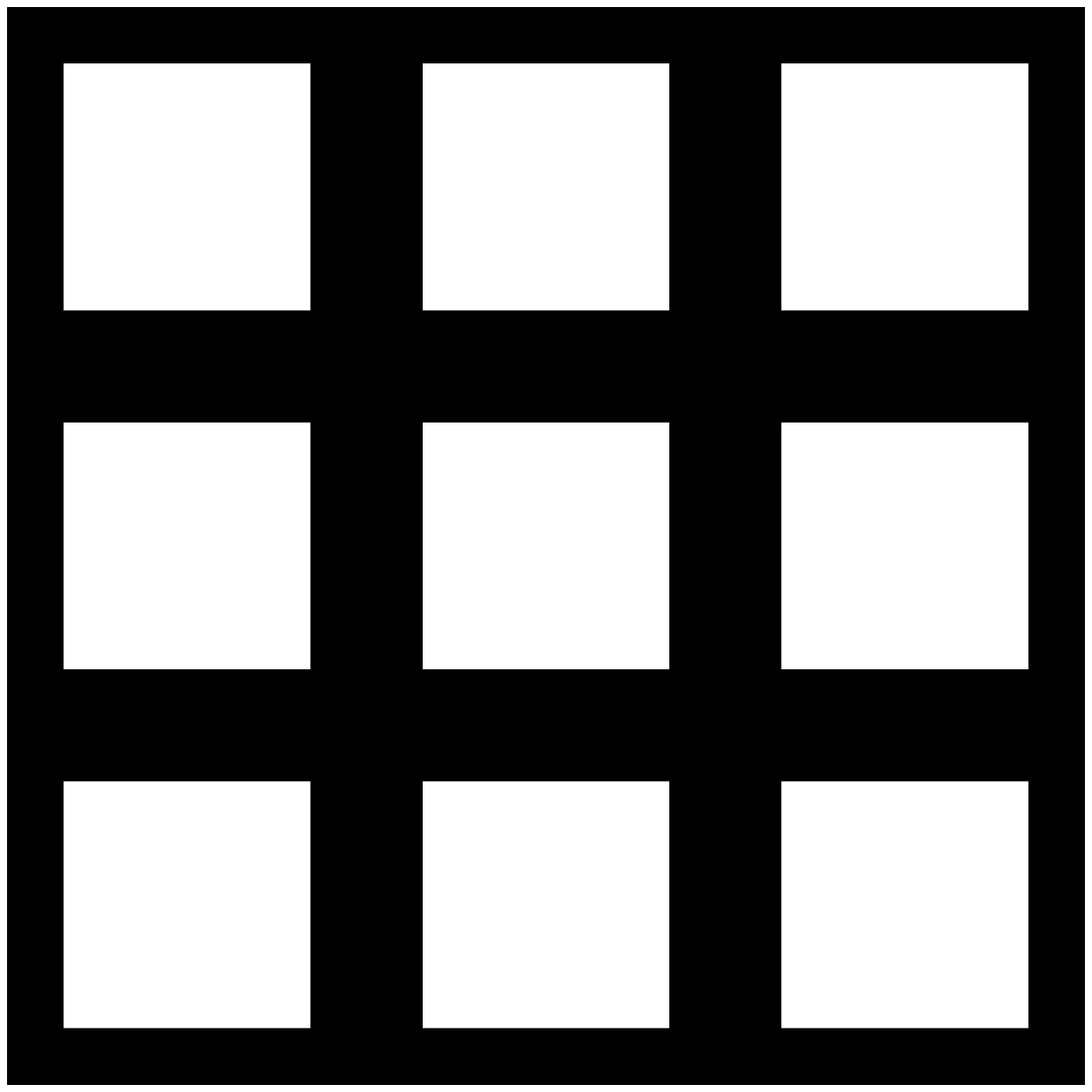}
  \includegraphics[width=32mm]{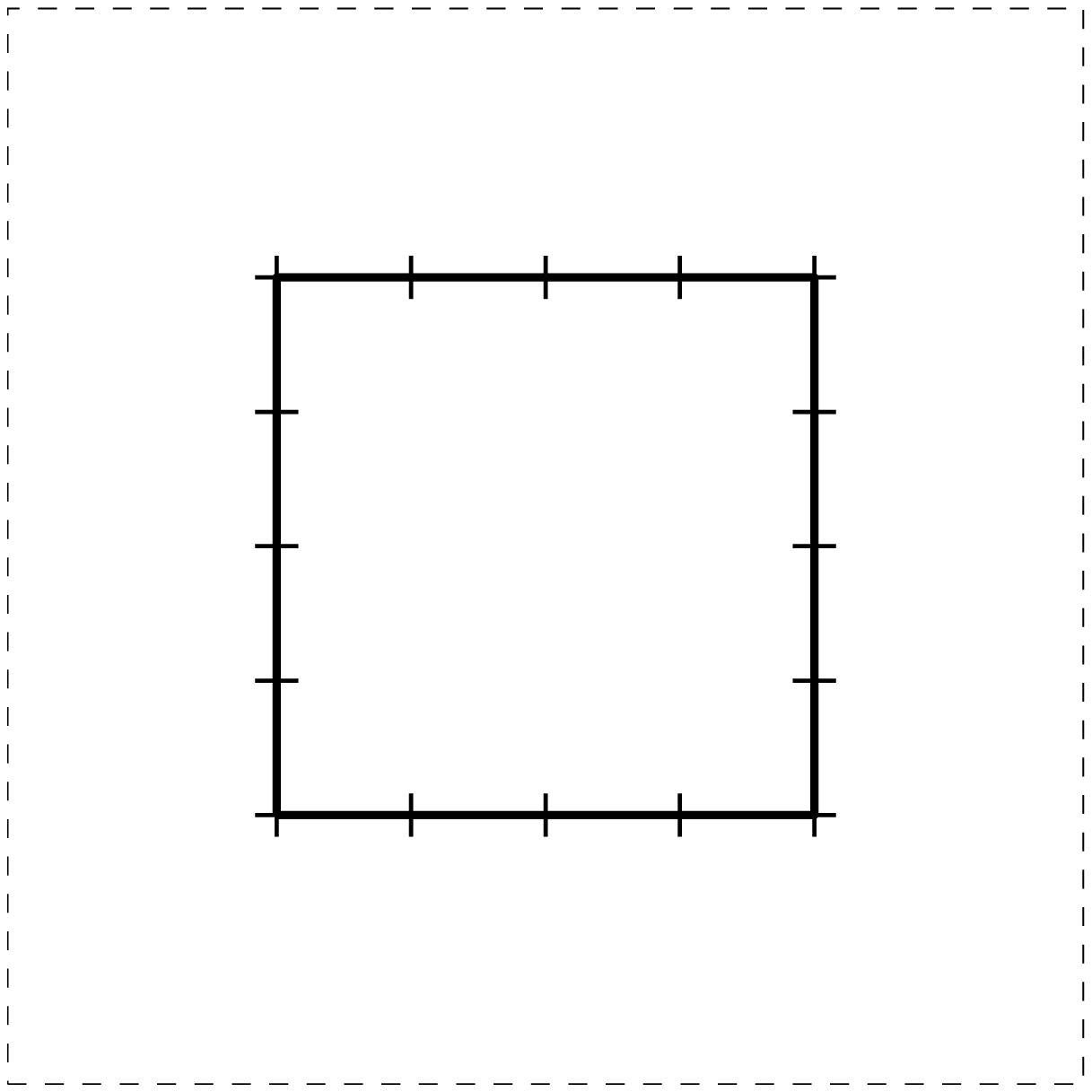}
  \hspace{2mm}
  \includegraphics[width=32mm]{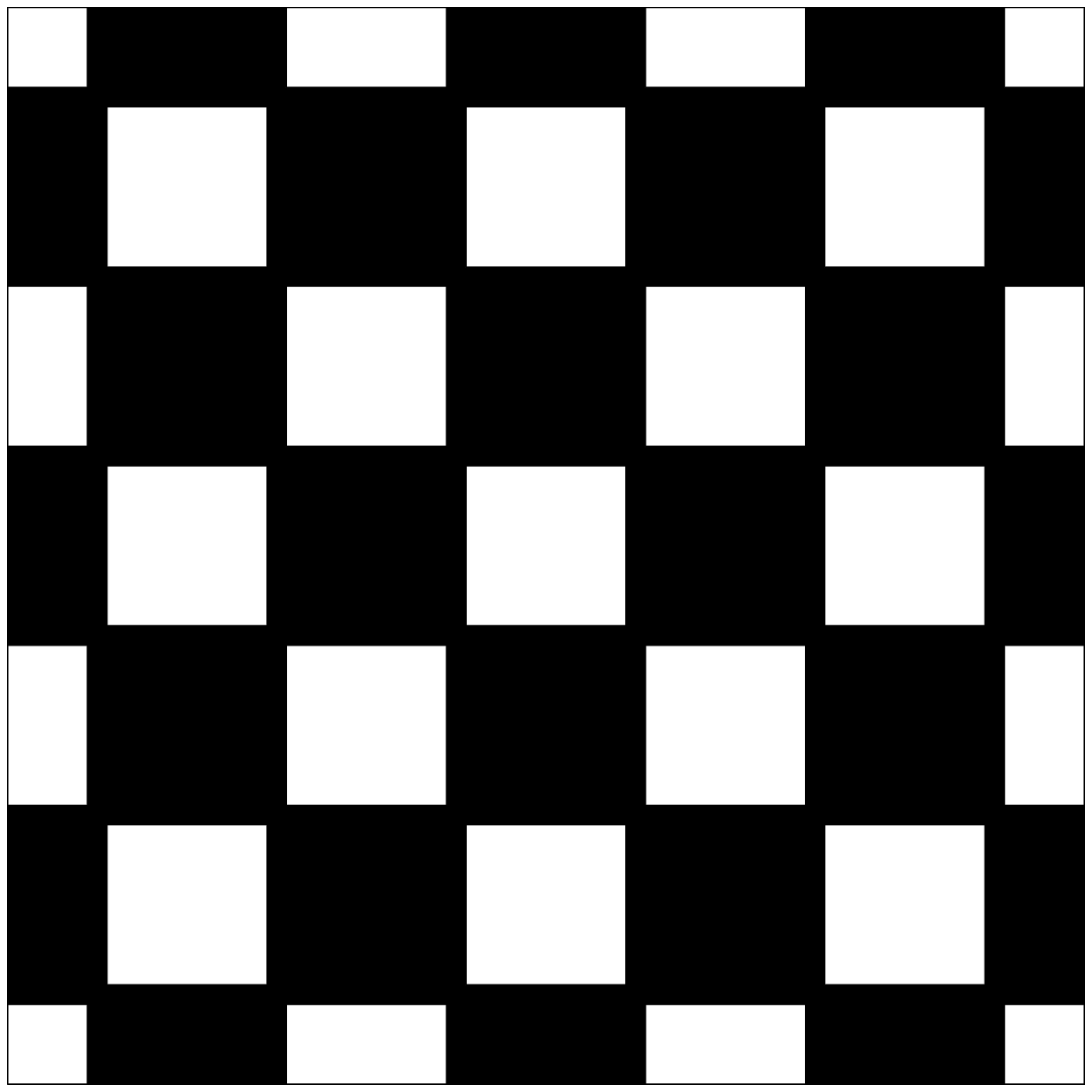}
  \includegraphics[width=32mm]{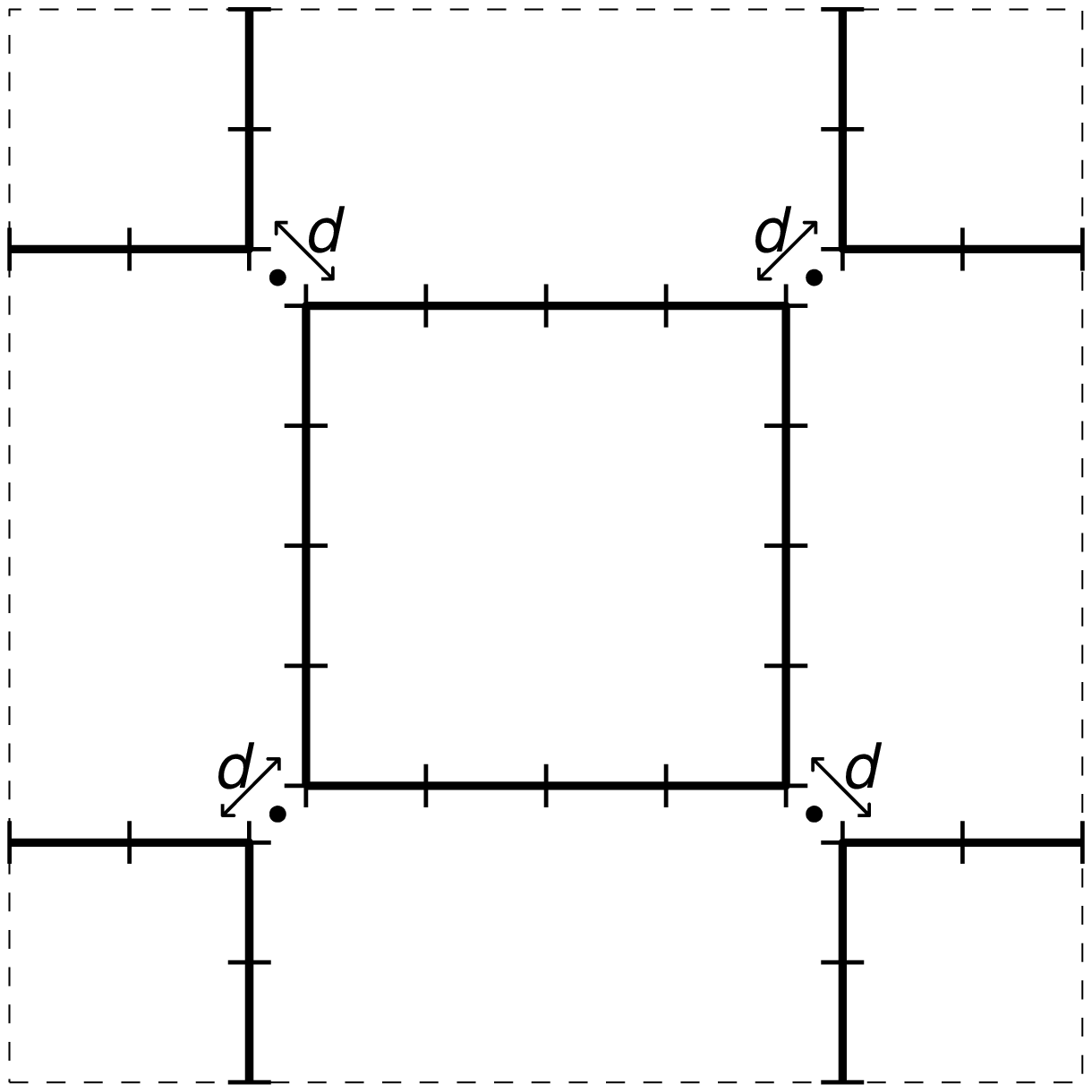}
\end{center}
\caption{\sf Left: a cutout of a square array of squares with 
  $p=0.5$ and a unit cell with a 16-panel coarse mesh on $\Gamma_0$.
  Right: a staggered array of squares with $p=0.4$ and a unit cell
  with a 32-panel coarse mesh on $\Gamma_0$. The vertex separation
  distance is $d$ and the dots indicate double-corner concentration
  points $\delta_k$.}
\label{fig:F1}
\end{figure}

\subsection{Ordered arrays of squares}
\label{sec:ordarr}

The two ordered arrays are made by placing squares with conductivity
$\sigma_2$ in a plane with conductivity $\sigma_1$. Fig.~\ref{fig:F1}
shows cutouts of unit cells. The orientation of $\Gamma$ is positive.
The arrays are overall isotropic, so $\sigma_{\ast}$ is independent of
$e$. The points in between neighboring corner vertices in the
staggered array are called {\it double-corner concentration points}
and denoted $\delta_k$.

The conductivity $\sigma_2$ may be complex valued while $\sigma_1$ is
assumed real. The special case of real valued and negative ratios
$\sigma_2/\sigma_1$ poses a particular challenge. The electrostatic
equation may not have a unique solution and this property is then
carried over to the integral equation. Sometimes $\sigma_{\ast}$,
viewed as a function of $\sigma_2$ with $\sigma_1$ held constant, has
a well defined limit which depends on whether $\sigma_2/\sigma_1$
approaches the negative real axis from above or from below in the
complex plane. Hetherington and Thorpe~\cite{Heth92} argue that such a
branch cut occurs for $\sigma_{\ast}$ of composites with right-angled
interfaces whenever $\sigma_2/\sigma_1\in[-3,-1/3]$. See also p.~378
of Milton~\cite{Milt02}. We shall capture the limit of $\sigma_{\ast}$
from above.

We follow standard practice for inclusion problems and represent
$U(z)$ as a continuous function which is a sum of a driving term and a
single-layer potential with density $\rho(z)$~\cite{Gree06}. Enforcing
continuity of the normal current across $\Gamma$ we arrive at the
integral equation
\begin{equation}
\rho(z)+\frac{\lambda(z)}{\pi}\int_{\Gamma}\rho(\tau)
\Im\left\{\frac{n_z\bar{n}_{\tau}\,{\rm d}\tau}{\tau-z}\right\}=
2\lambda(z)\Re\left\{\bar{e}n_z\right\}\,,\quad z\in\Gamma_0\,,
\label{eq:int1}
\end{equation}
where the `bar' symbol denotes complex conjugation. We observe
that~(\ref{eq:int1}) is a Fredholm integral equation of the second
kind with an integral operator which is compact away from the corner
vertices.

The parameter $\lambda(z)$ in~(\ref{eq:int1}) is independent of $z$.
Should the integral operator in~(\ref{eq:int1}) have been compact
everywhere, then, in a finite portion of the complex $\lambda$-plane,
there could exist a finite number of values $|\lambda|\ge 1$, called
eigenvalues of the equation, for which the solution $\rho(z)$ may not
be unique or may not even exist as a limit. See Sections 8 and 38 of
Mikhlin~\cite{Mikh64}. If, however, (\ref{eq:int1}) can be solved for
$\rho(z)$ and under the assumption that the inclusions do not overlap
the unit cell boundary, the effective conductivity can be computed
from
\begin{equation}
\sigma_{\ast}=\sigma_1+
\sigma_1\int_{\Gamma_0}\rho(z)\Re\left\{\bar{e}z\right\}\,{\rm d}|z|\,.
\label{eq:eff1}
\end{equation}

Depending on how the unit cell is chosen, the squares in the staggered
array may overlap the unit cell boundary. With the choice in
Fig.~\ref{fig:F1}, they certainly do. But since the layer density
$\rho(z)$ is periodic and identical on all squares one can
modify~(\ref{eq:eff1}) so that it integrates $\rho(z)$ twice on the
square at the center of the unit cell and ignores $\rho(z)$ on the
other squares.

\begin{figure}[t]
\begin{center}
  \includegraphics[width=41mm]{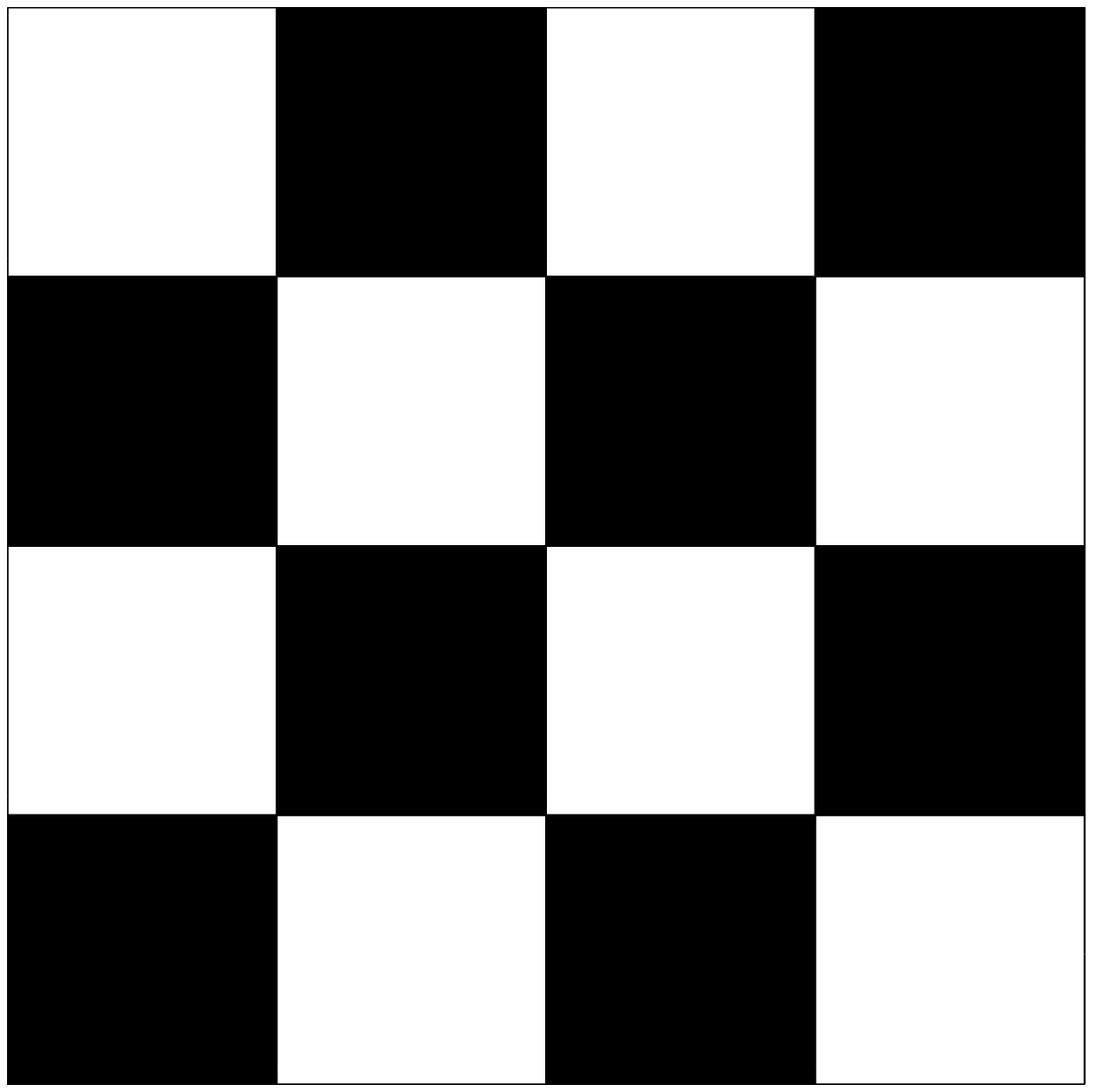}
  \hspace{2mm}
  \includegraphics[width=41mm]{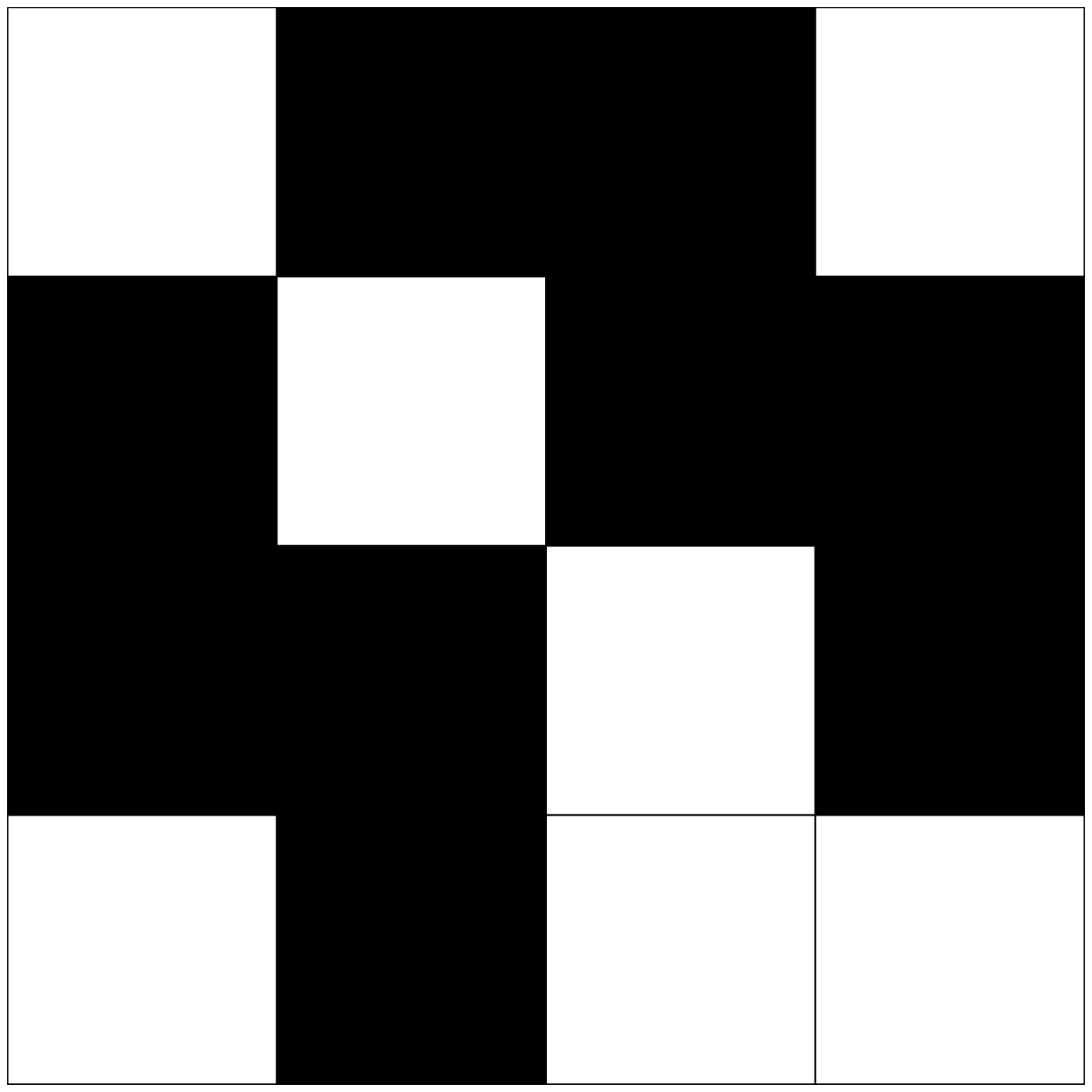}
  \hspace{2mm}
  \includegraphics[width=42mm]{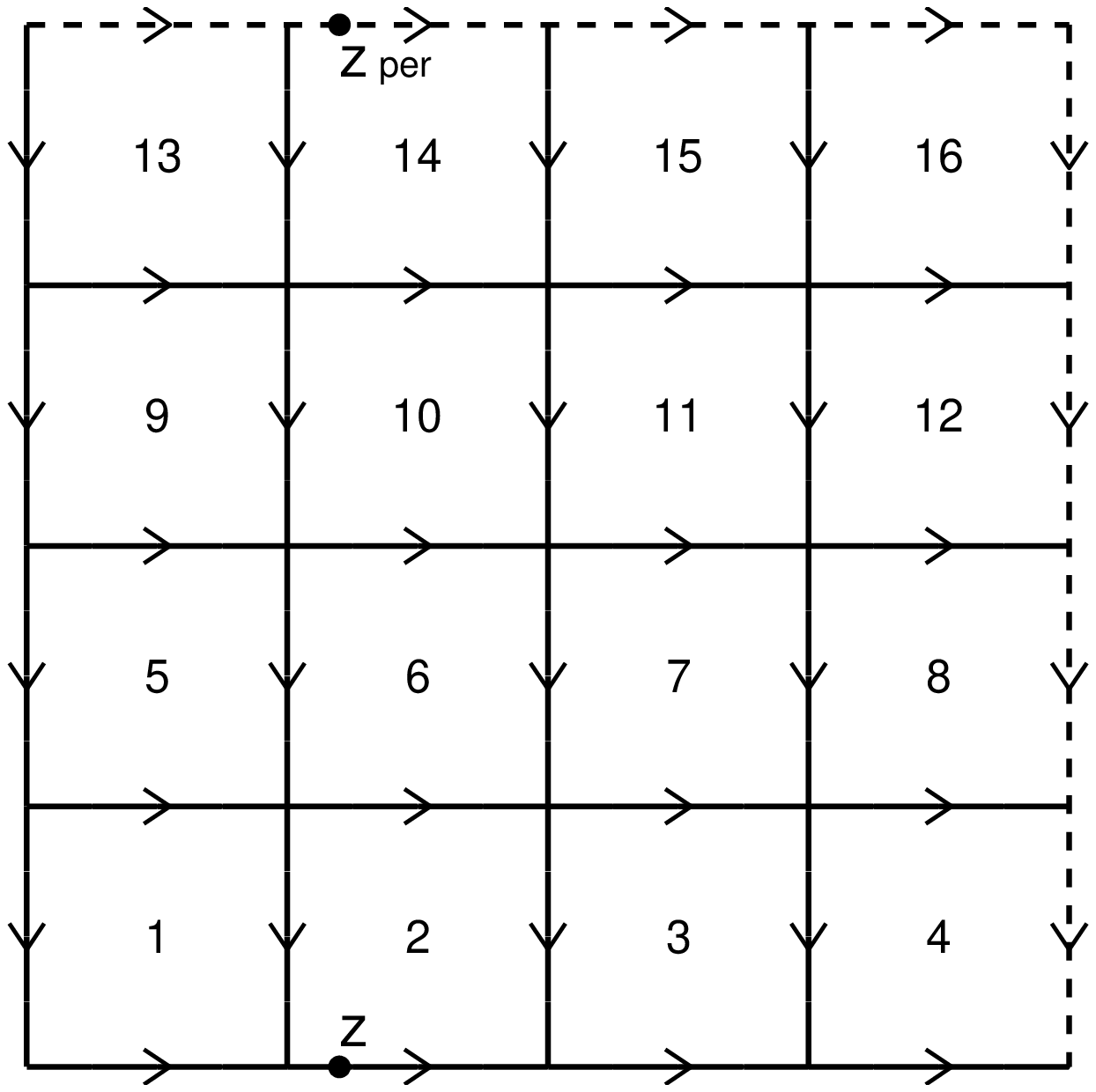}
\end{center}
\caption{\sf Left: a cutout of a two-component ordered checkerboard.
  Middle and right: A unit cell $D_0$ of a random checkerboard with
  $N_{\rm sq}=16$ squares. Orientation of $\Gamma_0$ (solid lines) and
  $L_0\setminus\Gamma_1$ (dashed lines) along with a point
  $z\in\Gamma_1$ and its periodic image $z_{\rm
    per}\in L_0\setminus\Gamma_1$.}
\label{fig:F2}
\end{figure}

\subsection{Checkerboards}

Fig.~\ref{fig:F2} shows checkerboards. The squares in $D$ have either
high conductivity $\sigma_2$ or low conductivity $\sigma_1$. Here
$\sigma_2$ and $\sigma_1$ are real so that $\sigma_2/\sigma_1>1$. The
challenge is to achieve linear complexity and high accuracy in
difficult situations.

The middle image of Fig.~\ref{fig:F2} is from a random checkerboard
with $N_{\rm sq}\!=\!16$. The right image indicates $\Gamma_0$ by
solid lines. The boundary of $D_0$ is denoted $L_0$ and
$\Gamma_1=\Gamma_0\cap L_0$. Note that some or all squares that meet
at a corner vertex $\gamma_k$ could have the same conductivity.
Vertices where two squares of conductivity $\sigma_1$ and two squares
of conductivity $\sigma_2$ meet diagonally, like in the left image of
Fig.~\ref{fig:F2}, will be referred to as {\it special corner
  vertices}.

An efficient Fredholm second kind integral equation for checkerboard
problems can be derived by applying Green's third identity to the
periodic function $U(z)-\Re\{\bar{e}z\}$. In terms of a transformed
potential $u(z)$, this double-layer type equation assumes the simple
form
\begin{equation}
u(z)-\frac{\lambda(z)}{\pi}\int_\Gamma
u(\tau)\Im\left\{\frac{{\rm d}\tau}{\tau-z}\right\}=
2\frac{c(z)}{b(z)}\Re\left\{\bar{e}l_0(z)\right\}\,,
\quad z \in \Gamma_0\,,
\label{eq:int2}
\end{equation}
where $l_0(z)$ is zero for $z\in\Gamma_0\setminus\Gamma_1$ and equal
to the vector difference of $z$ and its periodic image $z_{\rm per}\in
L_0\setminus\Gamma_1$ for $z\in\Gamma_1$, see Section~2.2
of~\cite{Hels11b}. The effective conductivity can be computed from
\begin{equation}
\sigma_{\ast}=
\int_{\Gamma_0} u(z)\Im\left\{\bar{e}\,{\rm d}z\right\}\,.
\label{eq:eff2}
\end{equation}
We observe that the integral operator in~(\ref{eq:int2}) is compact
away from the corner vertices.

\begin{figure}[t]
\begin{minipage}[b]{0.4\linewidth}
\begin{center}
\includegraphics[width=42mm]{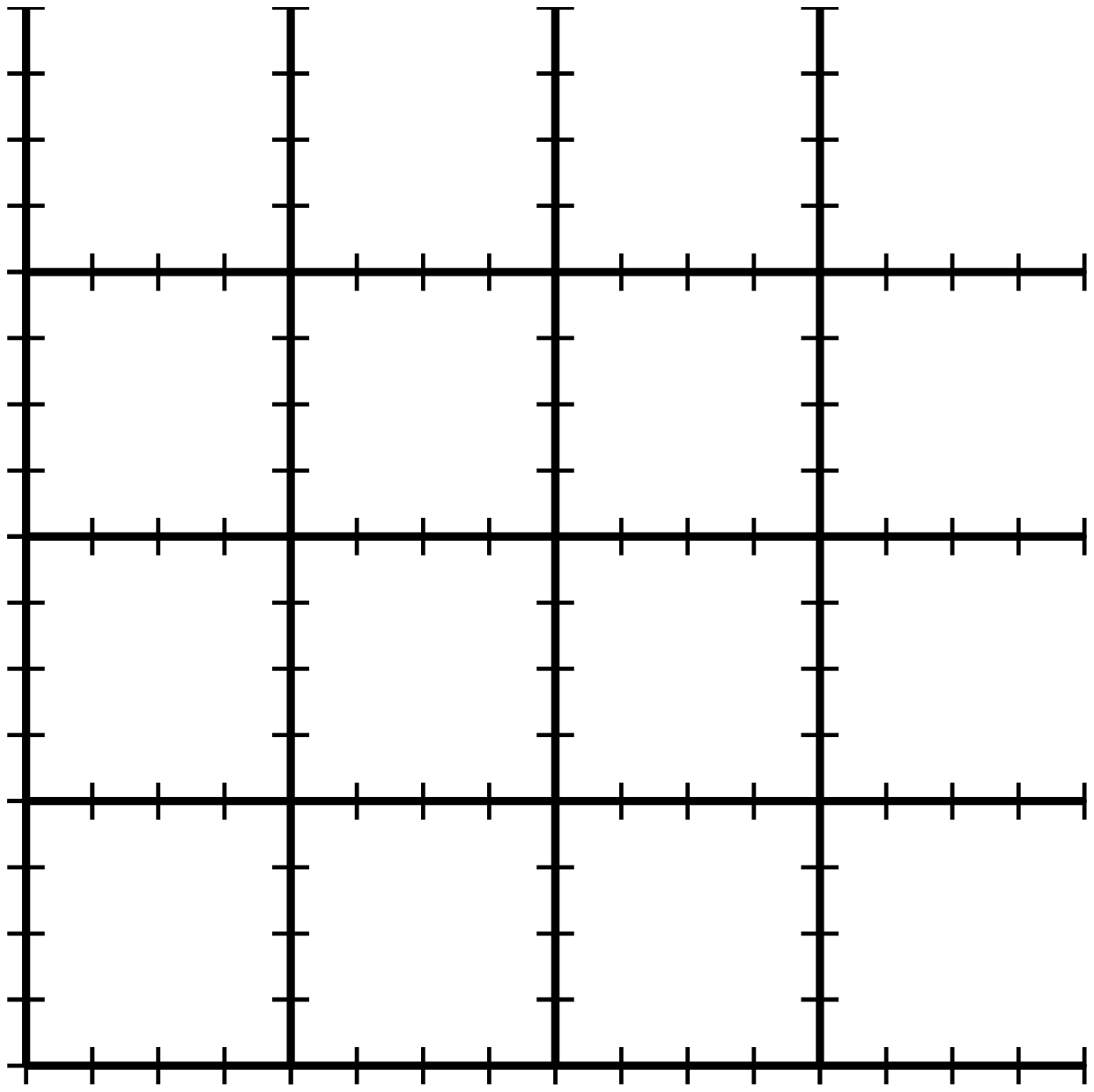}
\end{center}
\end{minipage}
\hspace{5mm}
\begin{minipage}[b]{0.5\linewidth}
\begin{center}
\includegraphics[height=25mm]{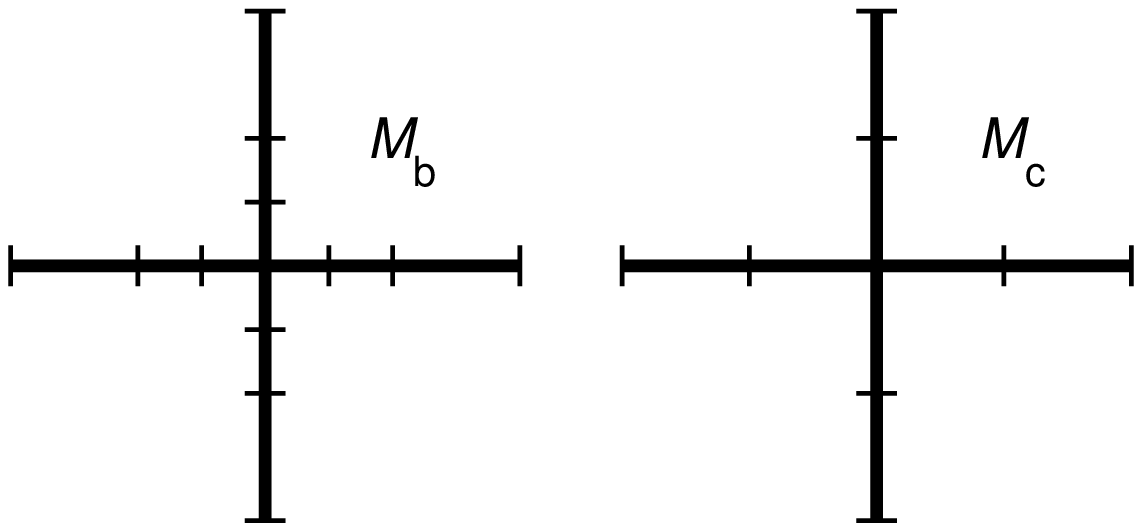}
\end{center}
\vspace{4.5mm}
\end{minipage}
\caption{\sf Left: a coarse mesh on $\Gamma_0$ for a checkerboard unit
  cell with $N_{\rm sq}=16$. There are four quadrature panels on each
  square side. Right: local meshes ${\cal M}_{\rm b}$ and ${\cal
    M}_{\rm c}$ centered around a corner vertex. There will be $192$
  discretization points on ${\cal M}_{\rm b}$ and $128$ points on
  ${\cal M}_{\rm c}$.}
\label{fig:checkmesh}
\end{figure}

\section{Discretization}
\label{sec:disc}

We discretize~(\ref{eq:int1}) and~(\ref{eq:int2}) using a Nystr{\"o}m
scheme based on composite polynomial interpolatory quadrature and a
parameterization $z(t)$ of $\Gamma_0$. {\it Coarse meshes} with four
quadrature panels per square side are constructed on $\Gamma_0$, see
Figs.~\ref{fig:F1} and~\ref{fig:checkmesh}. We also need {\it fine
  meshes} obtained from coarse meshes by subdividing panels
neighboring corner vertices $n_{\rm sub}$ times in a direction towards
the vertices.

The layer densities $\rho(z)$ and $u(z)$ in~(\ref{eq:int1})
and~(\ref{eq:int2}) are smooth on most quadrature panels. We choose
quadrature nodes and weights according to composite 16-point
Gauss--Legendre quadrature in parameter $t$ on such panels. This
quadrature has panelwise polynomial degree 31.

Panels neighboring corner vertices of ordered arrays of squares or
special corner vertices of checkerboards require special attention and
will be referred to as {\it special panels}. The layer densities
$\rho(z)$ and $u(z)$ may undergo rapid changes there. This is so
because of strong singularities that arise in $\nabla U(z)$. For
checkerboards, as $\sigma_2/\sigma_1\to\infty$, this field is barely
square integrable in $D_0$ and barely absolutely integrable on
$\Gamma_0$, see Section~2.3 of ~\cite{Hels11b}. See p.~378 of
Milton~\cite{Milt02} for a discussion of similar singularities that
arise at corner vertices as $\sigma_2/\sigma_1$ approaches values in
the range $[-3,-1/3]$.

Legendre nodes are not optimal for capturing the behavior of layer
densities on special panels. Rather, it pays off to bunch quadrature
nodes in a direction towards the vertices. An experimental
investigation, see Section~\ref{sec:accu}, shows that nodes
corresponding to zeros of the Jacobi polynomial
$P(x)_{16}^{(\alpha,\beta)}$ on the canonical interval $x\in[-1,1]$,
for certain $\alpha$ and $\beta$, are more efficient. For
checkerboards and with the corner vertex at a special panel's right
endpoint, we take $\alpha\!=\!(\sigma_1/\sigma_2)^{0.4}\!-\!1$ and
$\beta\!=\!0$. With the corner vertex at a special panel's left
endpoint, we take $\alpha\!=\!0$ and
$\beta\!=\!(\sigma_1/\sigma_2)^{0.4}\!-\!1$. For ordered arrays of
squares at negative contrast ratios we take $\alpha\!=\!10^{-6}-1$ and
$\beta\!=\!0$ or $\alpha\!=\!0$ and $\beta\!=\!10^{-6}-1$. The
corresponding quadrature weights are determined so that the panelwise
polynomial degree is 15.

A discretization in parameter $t$ on the coarse mesh of a checkerboard
gives $N_{\rm coa}\!=\!128N_{\rm sq}$ points $z_i\!=\!z(t_i)$ and the
same number of weights $w_i$. On the fine mesh there are $N_{\rm
  fin}\!=\!(128+64n_{\rm sub})N_{\rm sq}$ discretization points. The
square array of squares has $N_{\rm coa}\!=\!256$ and $N_{\rm
  fin}\!=\!256+64n_{\rm sub}$. The staggered array of squares has
$N_{\rm coa}\!=\!512$ and $N_{\rm fin}\!=\!512+128n_{\rm sub}$. We
collect quadrature weights on the diagonal of matrices ${\bf W}$ for
later use. The subscripts `coa' and 'fin' are used to indicate the
coarse mesh and the refined mesh, respectively.

\section{Short-range preconditioning}
\label{sec:short}

Consider now the Fredholm second kind integral
equations~(\ref{eq:int1}) and~(\ref{eq:int2}) in the general
form~(\ref{eq:gen0}) where $g(z)$ is piecewise smooth. Let $K(\tau,z)$
denote the kernel of $K$. Split $K(\tau,z)$ into two functions
\begin{equation}
K(\tau,z)=K^{\star}(\tau,z)+K^{\circ}(\tau,z)\,,
\label{eq:split}
\end{equation}
where $K^{\star}(\tau,z)$ is zero except for when $\tau$ and $z$
simultaneously lie in a neighborhood $\Gamma^{\star}_k$ centered
around a particular $\gamma_k$ or $\delta_k$. Then $K^{\circ}(\tau,z)$
is zero. The neighborhoods $\Gamma^{\star}_k$ cover four coarse panels
around $\gamma_k$ of a square array of squares and eight coarse panels
around $\delta_k$ of a staggered array of squares and around
$\gamma_k$ of a checkerboard. Compare Section 3.2 of~\cite{Hels11b}.

The kernel split~(\ref{eq:split}) corresponds to an operator split
$K=K^{\star}+K^{\circ}$ where $K^{\circ}$ is a compact operator. After
discretization~(\ref{eq:gen0}) assumes the form
\begin{equation}
\left({\bf I}+{\bf K}^{\star}+{\bf K}^{\circ}\right)\boldsymbol{\mu}={\bf g}\,,
\label{eq:gen1}
\end{equation}
where ${\bf I}$, ${\bf K}^{\star}$, and ${\bf K}^{\circ}$ are square
matrices and $\boldsymbol{\mu}$ and ${\bf g}$ are columns vectors.
Note that ${\bf K}^{\star}$ is sparse and block diagonal. The blocks
of ${\bf K}^{\star}_{\rm coa}$ corresponding to $\gamma_k$ of a square
array of squares have size $64\times 64$ while the blocks
corresponding to $\delta_k$ of a staggered array of squares and to
$\gamma_k$ of a checkerboard have size $128\times 128$.

The change of variables
\begin{equation}
\mu(z)=\left(I+K^{\star}\right)^{-1}\tilde{\mu}(z)
\end{equation}
makes~(\ref{eq:gen1}) read
\begin{equation}
\left({\bf I}+{\bf K}^{\circ}\left({\bf I}+{\bf K}^{\star}\right)^{-1}\right)
\tilde{\boldsymbol{\mu}}={\bf g}\,.
\label{eq:gen2}
\end{equation}
This right preconditioned equation corresponds to the discretization
of a Fredholm second kind equation with a composed compact operator
and the solution $\tilde{\boldsymbol{\mu}}$ is the discretization of a
piecewise smooth function. There should be no ill-conditioning
in~(\ref{eq:gen2}) due to mesh refinement close to corner vertices and
we can view $\left({\bf I}+{\bf K}^{\star}\right)^{-1}$ as a
short-range preconditioner for~(\ref{eq:gen1}). There will, however,
be ill-conditioning in~(\ref{eq:gen2}) for parameter values
$\lambda(z)$ that are very close to eigenvalues of~(\ref{eq:int1})
and~(\ref{eq:int2}).

\section{Compression of the preconditioned equation}
\label{sec:comp}

The matrix ${\bf K}^{\circ}$ and the right hand side ${\bf g}$
in~(\ref{eq:gen2}) can be accurately evaluated on a grid on the coarse
mesh. Only $\left({\bf I}+{\bf K}^{\star}\right)^{-1}$ needs a grid on
the refined mesh for its accurate evaluation. We introduce the
compressed weighted inverse
\begin{equation}
{\bf R}={\bf P}^T_W
\left({\bf I}_{\rm fin}+{\bf K}_{\rm fin}^{\star}\right)^{-1}{\bf P}\,.
\label{eq:R0}
\end{equation}
Here ${\bf P}$ is a prolongation operator from the coarse grid to the
fine grid, ${\bf P}_W={\bf W}_{\rm fin}{\bf P}{\bf W}_{\rm coa}^{-1}$
is a weighted prolongation operator, see Section~5 of~\cite{Hels09b}.
Furthermore, the block-diagonal $N_{\rm coa}\times N_{\rm coa}$ matrix
${\bf P}_W^T{\bf P}$, where superscript $T$ denotes the transpose, has
the property
\begin{equation}
{\bf P}_W^T{\bf P}={\bf I}\,.
\label{eq:iden}
\end{equation}
Strictly speaking, the relation~(\ref{eq:iden}) does not hold exactly
for matrix blocks corresponding to special panels. It holds, however,
also for these blocks that
\begin{equation}
{\bf f}_i{\bf W}_{\rm coa}{\bf P}_W^T{\bf P}{\bf f}_j=
{\bf f}_i{\bf W}_{\rm coa}{\bf f}_j\,,
\end{equation}
where ${\bf f}_i$ and ${\bf f}_j$ are discretizations of piecewise
polynomials on the coarse grid of degree $i$ and $j$ and $i+j\le 15$.
One can say that~(\ref{eq:iden}) holds to the same polynomial degree
as the overall quadrature holds.

With~(\ref{eq:R0}), equation~(\ref{eq:gen2}) assumes the form
\begin{equation}
\left({\bf I}_{\rm coa}+{\bf K}_{\rm coa}^{\circ}{\bf R}\right)
\tilde{\boldsymbol{\mu}}_{\rm coa}={\bf g}_{\rm coa}\,.
\label{eq:comp1}
\end{equation}
The single-layer equation~(\ref{eq:int1}) will be used in this form in
the numerical examples of Section~\ref{sec:numex}. In terms of the new
discrete density $\hat{\boldsymbol{\mu}}_{\rm coa}= {\bf
  R}\tilde{\boldsymbol{\mu}}_{\rm coa}$ one can also
write~(\ref{eq:comp1}) in left preconditioned form
\begin{equation}
\left({\bf I}_{\rm coa}+{\bf R}{\bf K}_{\rm coa}^{\circ}\right)
\hat{\boldsymbol{\mu}}_{\rm coa}={\bf R}{\bf g}_{\rm coa}\,.
\label{eq:comp2}
\end{equation}
The double-layer equation~(\ref{eq:int2}) will be used in this form in
the more elaborate scheme for complicated unit cells developed in
Section~\ref{sec:long}.

Functionals on $\mu(z)$ of the type
\begin{equation}
\int f(z)\mu(z)\,{\rm d}z=\int f(z(t))\mu(z(t))\,z'(t)\,{\rm d}t\,,
\label{eq:gen3}
\end{equation}
where $f(z)$ is a piecewise smooth function, assume the discretized
form
\begin{equation}
{\bf f}_{\rm coa}^T{\bf Z}_{\rm coa}\hat{\boldsymbol{\mu}}_{\rm coa}\,,
\label{eq:gen3D}
\end{equation}
where ${\bf f}$ is a column vector and ${\bf Z}$ is a matrix
containing discrete values $z'_i\!=\!z'(t_i)$ multiplied with weights
$w_i$ on the diagonal.

\begin{figure}[t]
\begin{center}
\includegraphics[height=50mm]{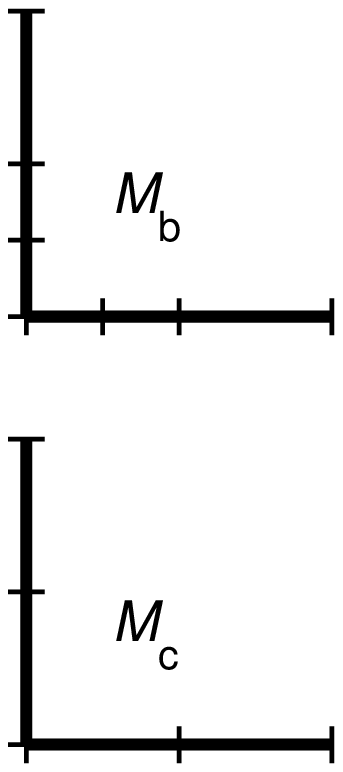}
\hspace{10mm}
\includegraphics[height=50mm]{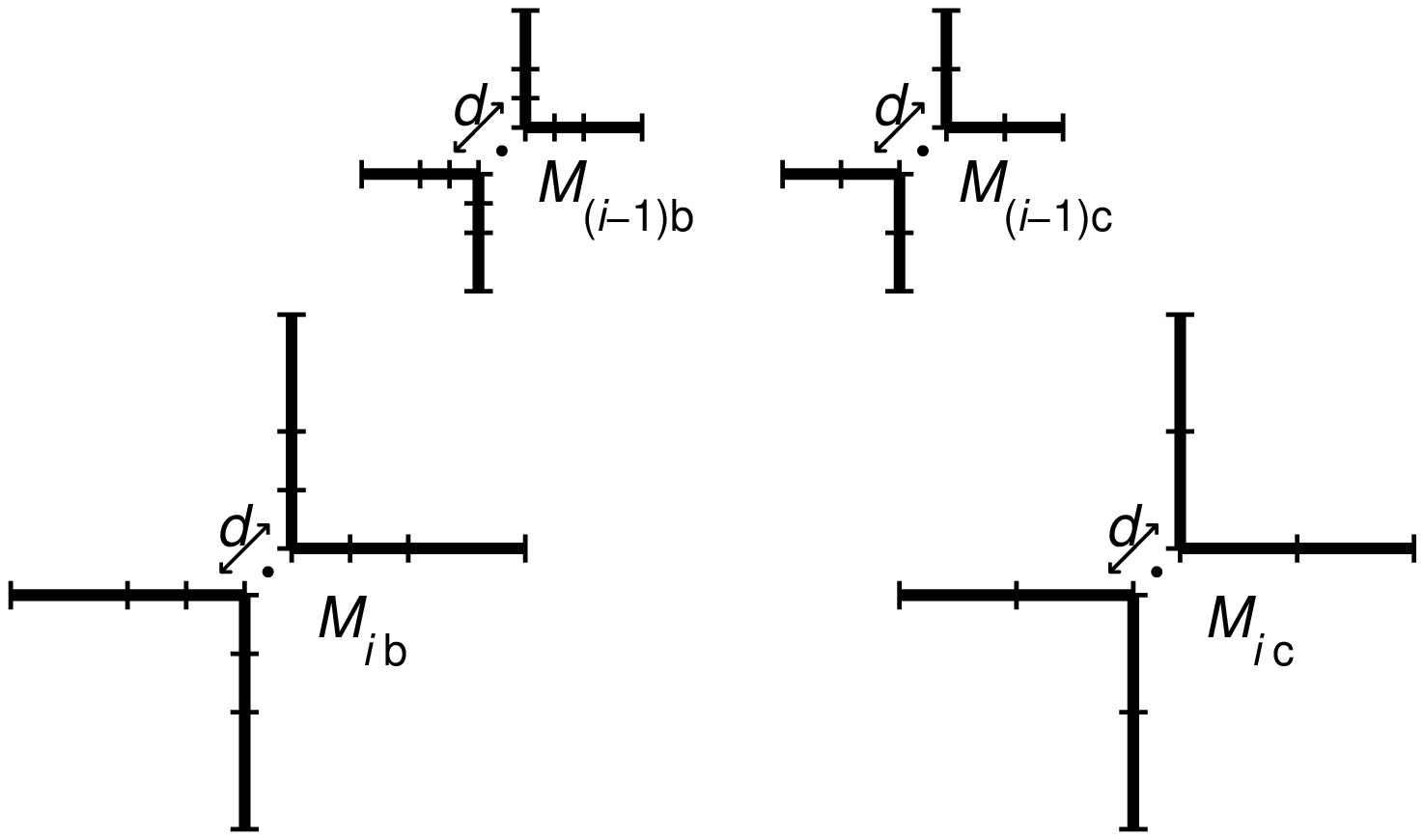}
\end{center}
\caption{\sf Local meshes close to corner vertices. Left: the square 
  array. Right: the staggered array with meshes centered around a
  double-corner concentration point $\delta_k$. The vertex separation
  distance is $d$. Meshes with index $i\!=\!n_{\rm rec}$ have the
  largest panels and ${\cal M}_{n{\rm c}}$ coincides with the coarse
  mesh on $\Gamma_0$ in a neighborhood of $\delta_k$, see the
  rightmost image of Fig.~\ref{fig:F1}. The panels on meshes with
  index $i\!-\!1$ are half the size of those on meshes with index
  $i$.}
\label{fig:stagmesh}
\end{figure}

\section{Recursive construction of ${\bf R}$}
\label{sec:recur}

The compressed inverse ${\bf R}$ has the same block diagonal structure
as ${\bf K}^{\star}_{\rm coa}$, see Section~\ref{sec:short}. Its
construction from the definition~(\ref{eq:R0}) is costly when the
refined mesh has many panels. Actually, the number of subdivisions
needed to reach a given accuracy may grow without bounds due to the
singularities in $\mu(z)$ that arise as $\sigma_2/\sigma_1$ approaches
certain values, see Section~\ref{sec:disc}.

Fortunately, the construction of each block ${\bf R}_k$ of ${\bf R}$,
associated with a corner vertex $\gamma_k$ or with a double-corner
concentration point $\delta_k$, can be greatly sped up and also
stabilized via a recursion. This recursion uses grids on local meshes
centered around a $\gamma_k$ or a $\delta_k$, see
Figs.~\ref{fig:checkmesh} and~\ref{fig:stagmesh}.

\subsection{General recursion}

The staggered array of squares needs the recursion in the general form
\begin{equation}
{\bf R}_{ik}={\bf P}^T_{W\rm{bc}}
\left(
\mathbb{F}\{{\bf R}_{(i-1)k}^{-1}\}
+{\bf I}_{\rm b}^{\circ}+{\bf K}_{i{\rm b}k}^{\circ}
\right)^{-1}{\bf P}_{\rm{bc}}\,,\quad i=1,\ldots,n_{\rm rec}\,,
\label{eq:genrec}
\end{equation}
where the number of recursion steps $n_{\rm rec}$ corresponds to
$n_{\rm sub}$ of the refined mesh and where ${\bf R}_{ik}={\bf R}_{k}$
for $i\!=\!n_{\rm rec}$. See Section~6 of~\cite{Hels09b}. The weighted
and unweighted prolongation operators ${\bf P}_{W\rm{bc}}$ and ${\bf
  P}_{\rm{bc}}$ act from a 128-point grid on a local mesh ${\cal
  M}_{i{\rm c}}$ to a 192-point grid on a local mesh ${\cal M}_{i{\rm
    b}}$, see the right image of Fig.~\ref{fig:stagmesh}.  The
superscript '$\circ$' in~(\ref{eq:genrec}) has a meaning which can be
explained by considering the discretization of $K$ on a 192-point grid
on ${\cal M}_{i{\rm b}}$ and on a 128-point grid on ${\cal
  M}_{(i-1){\rm c}}$.  Let the resulting matrices be ${\bf K}_{i{\rm
    b}k}$ and ${\bf K}_{(i-1){\rm c}k}$. Now ${\bf K}_{i{\rm
    b}k}^{\circ}$ is the $192\times 192$ matrix which results from
zeroing all entries of ${\bf K}_{i{\rm b}k}$ that also are contained
in the $128\times 128$ matrix ${\bf K}_{(i-1){\rm c}k}$. The operator
$\mathbb{F}\{\cdot\}$ expands an $128\times 128$ matrix into an
$192\times 192$ matrix by zero-padding in such a way that
$\mathbb{F}\{{\bf K}_{(i-1){\rm c}k}\}+{\bf K}_{i{\rm
    b}k}^{\circ}={\bf K}_{i{\rm b}k}$.

\subsection{Fixed-point iteration and Newton's method}

The recursion~(\ref{eq:genrec}) can be simplified for square arrays of
squares and for checkerboards thanks to scale invariance of the
integrals in~(\ref{eq:int1}) and~(\ref{eq:int2}). The local meshes
${\cal M}_{i{\rm b}}$ and ${\cal M}_{i{\rm c}}$ look the same at all
recursion steps and the index $i$ can be dropped, see the right image
of Fig.~\ref{fig:checkmesh} and the left image of
Fig.~\ref{fig:stagmesh}. The recursion~(\ref{eq:genrec}) assumes the
form of a fixed-point iteration
\begin{equation}
{\bf R}_{ik}={\bf P}^T_{W\rm{bc}}
\left(
\mathbb{F}\{{\bf R}_{(i-1)k}^{-1}\}
+{\bf I}_{\rm b}^{\circ}+{\bf K}_{{\rm b}k}^{\circ}
\right)^{-1}{\bf P}_{\rm{bc}}\,,\quad i=1,\ldots,n_{\rm rec}\,,
\label{eq:fixed}
\end{equation}
which for $n_{\rm rec}\to\infty$ can be cast as a non-linear matrix
equation
\begin{equation}
{\bf G}({\bf R}_k)\equiv
{\bf P}^T_{W\rm{bc}}{\bf A}({\bf R}_k){\bf P}_{\rm{bc}}-{\bf R}_k=0\,,
\label{eq:Newton}
\end{equation}
where
\begin{equation}
{\bf A}({\bf R}_k)=\left(\mathbb{F}\{{\bf R}_k^{-1}\}
+{\bf I}_{\rm b}^{\circ}+{\bf K}_{{\rm b}k}^{\circ}\right)^{-1}\,,
\end{equation}
see Sections~3.2 and 3.3 of~\cite{Hels11b}. The non-linear
equation~(\ref{eq:Newton}), in turn, can be solved for ${\bf R}_k$
with a variant of Newton's method. Let ${\bf X}$ be a matrix-valued
perturbation of ${\bf R}_k$ and expand ${\bf G}({\bf R}_k\!+\!{\bf
  X})\!=\!0$ to first order in ${\bf X}$. This gives a Sylvester-type
matrix equation
\begin{equation}
{\bf X}-{\bf P}^T_{W\rm{bc}}{\bf A}({\bf R}_k)\mathbb{F}\{{\bf R}_k^{-1}
{\bf X}{\bf R}_k^{-1}\}{\bf A}({\bf R}_k){\bf P}_{\rm{bc}}={\bf G}({\bf R}_k)
\label{eq:Sylvester}
\end{equation}
for the Newton update ${\bf X}$. One can use the {\sc Matlab} built-in
function {\tt dlyap} for~(\ref{eq:Sylvester}), but GMRES~\cite{Saad86}
gives a smaller residual and we use that method.

\begin{figure}[t]
\begin{center}
  \includegraphics[height=69mm]{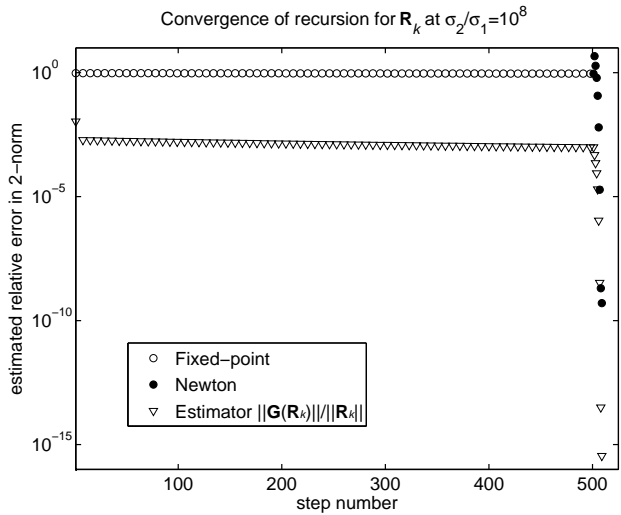}
\end{center}
\caption{\sf Convergence of ${\bf R}_k$ associated with a $\gamma_k$
  of an ordered checkerboard with $\sigma_2/\sigma_1=10^8$. The 500
  fixed-point iteration steps in~(\ref{eq:fixed}) are followed by 9
  Newton steps for~(\ref{eq:Newton}).}
\label{fig:recconv}
\end{figure}

\subsection{Initialization, number of recursion steps, and homotopy}

The recursion~(\ref{eq:genrec}), the fixed-point
iteration~(\ref{eq:fixed}), and Newton's method for~(\ref{eq:Newton})
need to be initialized and $n_{\rm rec}$ has to be decided
in~(\ref{eq:genrec}) and in~(\ref{eq:fixed}). The three types of
domains call for different strategies.

Random checkerboards at high contrast ratios are the easiest to deal
with. Here we first use the fixed-point iteration~(\ref{eq:fixed}),
initialized with
\begin{equation}
\mathbb{F}\{{\bf R}_{0k}^{-1}\}=
{\bf I}_{\rm c}+{\bf K}_{{\rm c}k}\,,
\label{eq:Ristart}
\end{equation}
where subscript `c' refers to a discretization on mesh ${\cal M}_{\rm
  c}$ in Fig.~\ref{fig:checkmesh}. Compare eq.~(24) of~\cite{Hels11b}.
The iterations are stopped when either $||{\bf G}({\bf R}_k)||/||{\bf
  R}_k||$ is smaller than $10\epsilon_{\rm mach}$ in Frobenius norm or
a number of 500 iterations is reached. The final fixed-point iterate
is then used as initial guess in Newton's method
for~(\ref{eq:Newton}). The Newton iterations are stopped when either
$||{\bf G}({\bf R}_k)||/||{\bf R}_k||$ is smaller than
$100\epsilon_{\rm mach}$ or a maximum number of 15 iterations is
reached. Fig.~\ref{fig:recconv} illustrates this strategy for an ${\bf
  R}_k$ associated with a $\gamma_k$ of an ordered checkerboard at
$\sigma_2/\sigma_1=10^8$. One can see that the convergence of the
fixed-point iteration is very slow. About $2\cdot 10^5$ steps,
corresponding to the same number of subdivisions of the fine mesh,
would be needed for full convergence if only the fixed-point iteration
was used. This clearly shows the power of Newton iterations and
explains why methods relying solely on mesh refinement run into great
difficulties on these types of domains. There are only 16 possible
corner configurations in a random checkerboard, corresponding to 16
distinct blocks of ${\bf R}$. Therefore, the time- and storage
requirements for computing ${\bf R}$ are negligible for large unit
cells.

Square arrays of squares at negative contrast ratios are more
difficult to treat. This is so since the solution to the electrostatic
equation may only exist as a limit for $\sigma_2/\sigma_1$ approaching
the negative real axis, see Section~\ref{sec:ordarr}. Again we first
use the fixed-point iteration~(\ref{eq:fixed}), initialized as
in~(\ref{eq:Ristart}) but with $\sigma_2$ (which enters into ${\bf
  K}_{{\rm b}k}^{\circ}$) multiplied with a constant $q=1-0.01{\rm
  i}$. Again the final fixed-point iterate is used as initial guess in
Newton's method for~(\ref{eq:Newton}). Now, however, we use a homotopy
method and at each Newton step we reduce the imaginary part of $q$
with a factor of ten. After 14 such iterations $q$ is set to unity and
an additional maximum of 15 Newton iterations are performed. In this
way the final expression for ${\bf R}_k$ may be complex valued even
though the last few matrices ${\bf K}_{{\rm b}k}^{\circ}$, fed
into~(\ref{eq:Newton}), are purely real.

Staggered arrays of squares at negative contrast ratios are the most
intricate. Here the choice of $n_{\rm rec}$ and of initializer
in~(\ref{eq:genrec}) are very important. We choose $n_{\rm rec}$ large
enough so that the vertex separation distance $d$, see
Fig.~\ref{fig:stagmesh}, at the first recursion step ($i\!=\!1$) is at
least $10^{16}$ times larger than the part of $\Gamma^{\star}_k$
covered by the mesh ${\cal M}_{1b}$. In this way the interaction
between the two connected parts of ${\cal M}_{1b}$ is negligible. The
initializer ${\bf R}_{0k}$ is then chosen as a compressed inverse
computed using the homotopy method just described for the ${\bf R}_k$
of the square array of squares, neglecting the interaction between the
connected parts of ${\cal M}_{1b}$.

\section{Long-range preconditioning}
\label{sec:long}

As the number of squares in a unit cell grows, the problem of
computing $\sigma_{\ast}$ gets harder, see Section~\ref{sec:motif}.
This section improves on Section~4 of~\cite{Hels11b} and describes a
long-range preconditioner for~(\ref{eq:int2}) which cures these
problems. The main idea is to split the unknown layer density into two
parts and capture all long-range interaction in a matrix ${\bf S}$,
which can be rapidly inverted and used in a right-preconditioner. In
combination with the short-range preconditioner ${\bf R}$ of
Section~\ref{sec:comp}, applied from the left, this results in a
scheme whose computational cost for checkerboards with large random
unit cells at high contrast ratios is almost linear in $N_{\rm sq}$.

\subsection{An expanded equation}

Each square in $D_0$ has a boundary consisting of four straight
segments, see Fig.~\ref{fig:F2}. Introduce piecewise constant local
basis functions $s_k(z)$, $k\!=\!1,\ldots,N_{\rm sq}$, on
$\Gamma_0\cup L_0$ such that $s_k(z)\!=\!1$ when $z$ lies on a
boundary part of square $k$ with positive orientation, $s_k(z)\!=\!-1$
when $z$ lies on a boundary part with negative orientation, and
$s_k(z)\!=\!0$ otherwise.

Following Section~4.1 of~\cite{Hels11b}, we take~(\ref{eq:int2}) in
the general form~(\ref{eq:gen0}) and expand it into the system
\begin{equation}
\left(I+K\right)\mu_0(z)+\sum_{k=1}^{N_{\rm sq}-1}a_k
\left(s_k(z)-\lambda(z)|s_k(z)|+\frac{2\lambda(z)}{N_{\rm sq}}\right)=g(z)\,.
\label{eq:int2b}
\end{equation}
\begin{equation}
\int s_k(z)\mu_0(z)\,{\rm d}|z|=0\,, \quad k=1,\ldots,N_{\rm sq}-1\,,
\label{eq:constr}
\end{equation}
where $\mu_0(z)$ mimics the rapidly varying behavior of $\mu(z)$ and
$a_k$ are unknown coefficients.

Discretization of~(\ref{eq:int2b}) and~(\ref{eq:constr}) together with
left preconditioned compression, compare~(\ref{eq:comp2}), results in
the linear system
\begin{equation}
\left({\bf I}_{\rm coa}+{\bf R}{\bf K}_{\rm coa}^{\circ}\right)
\hat{\boldsymbol{\mu}}_{0{\rm coa}}
+{\bf R}\left({\bf B}_1-\boldsymbol{\Lambda}_{\rm coa}|{\bf B}_1|
 +\boldsymbol{\lambda}_{\rm coa}{\bf u}^T\right){\bf a}
={\bf R}{\bf g}_{\rm coa}\,,
\label{eq:int2bD}
\end{equation}
\begin{equation}
{\bf B}_1^T|{\bf Z}_{\rm coa}|
\hat{\boldsymbol{\mu}}_{0{\rm coa}}={\bf 0}\,.
\label{eq:constrD}
\end{equation}
Here ${\bf B}_1$ is a $N_{\rm coa}\times (N_{\rm sq}\!-\!1)$ matrix
whose $k$th column is the discretization of $s_k(z)$,
$\boldsymbol{\lambda}_{\rm coa}$ is a column vector whose $N_{\rm
  coa}$ entries is the discretization of $\lambda(z)$,
$\boldsymbol{\Lambda}_{\rm coa}$ is matrix containing
$\boldsymbol{\lambda}_{\rm coa}$ on the diagonal, ${\bf u}$ is a
column vector with $N_{\rm sq}\!-\!1$ entries all equal to $2/N_{\rm
  sq}$, ${\bf a}$ is a column vector containing the $N_{\rm sq}\!-\!1$
coefficients $a_k$, and vertical bars denote entrywise absolute value.

The effective conductivity~(\ref{eq:eff2}) can be computed from
\begin{equation}
\sigma_{\ast}=\Im\left\{
\bar{{\bf e}}_{\rm coa}^T{\bf Z}_{\rm coa}\left(
\hat{\boldsymbol{\mu}}_{0{\rm coa}}+{\bf B}_1{\bf a}\right)
\right\}\,,
\label{eq:eff2D}
\end{equation}
once~(\ref{eq:int2bD}) and~(\ref{eq:constrD}) is solved.
Compare~(\ref{eq:gen3D}).

\subsection{An important simplification}

The definition of $s_k(z)$ together with Cauchy's integral theorem
implies that
\begin{equation}
\bar{{\bf e}}_{\rm coa}^T{\bf Z}_{\rm coa}{\bf B}_1={\bf 0}\,.
\end{equation}
As a consequence, the second term within parenthesis
in~(\ref{eq:eff2D}) does not contribute to $\sigma_{\ast}$ and can
be omitted.

The fact that the coefficients $a_k$ are not needed
in~(\ref{eq:eff2D}) opens up for another, more important,
simplification. With the change of variables
\begin{equation}
b_k=a_k+\frac{\left(\sigma(k)-\sigma(N_{\rm sq})\right)}{\sigma(N_{\rm sq})}
\frac{\sum_{i=1}^{N_{\rm sq}\!-\!1}a_i}{N_{\rm sq}}\,,
\end{equation}
where $\sigma(k)$ denotes the conductivity of square $k$, the expanded
equation~(\ref{eq:int2b}) assumes the simpler form
\begin{equation}
\left(I+K\right)\mu_0(z)+\sum_{k=1}^{N_{\rm sq}-1}b_k
\left(s_k(z)-\lambda(z)|s_k(z)|\right)=g(z)
\label{eq:int2c}
\end{equation}
and~(\ref{eq:int2bD}) reduces to
\begin{equation}
\left({\bf I}_{\rm coa}+{\bf R}{\bf K}_{\rm coa}^{\circ}\right)
\hat{\boldsymbol{\mu}}_{0{\rm coa}}
+{\bf R}\left({\bf B}_1-\boldsymbol{\Lambda}_{\rm coa}|{\bf B}_1|\right){\bf b}
={\bf R}{\bf g}_{\rm coa}\,.
\label{eq:int2bE}
\end{equation}

\subsection{A Schur complement style preconditioner}

The system~(\ref{eq:int2bE}) and~(\ref{eq:constrD}) can be written in
partitioned form
\begin{equation}
\left[ \begin{array}{cc}
{\bf I}+{\bf R}{\bf K}^{\circ} & {\bf B} \\
{\bf C}                        & {\bf 0}
\end{array} \right]
\left[\begin{array}{c}
 \hat{\boldsymbol{\mu}}_0 \\ 
 {\bf b} 
      \end{array}\right]
= \left[ \begin{array}{c}
{\bf R}{\bf g} \\ 0 \end{array} \right]\,,
\label{eq:disc2}
\end{equation}
where subscripts `coa' are omitted and
\begin{align}
{\bf B}&={\bf R}\left({\bf B}_1-\boldsymbol{\Lambda}_{\rm coa}|{\bf B}_1|
\right)\,,
\label{eq:B}\\
{\bf C}&={\bf B}_1^T|{\bf Z}_{\rm coa}|\,.
\label{eq:C}
\end{align}

The change of variables
\begin{equation}
\left[\begin{array}{c}
 \hat{\boldsymbol{\mu}}_0 \\ 
 {\bf b} 
      \end{array}\right]
=
\left[ \begin{array}{cc}
{\bf I} & {\bf B} \\
{\bf C} & {\bf 0}
\end{array} \right]^{-1}
\left[\begin{array}{c}
 \boldsymbol{\omega} \\ 
 {\bf c}
      \end{array}\right]
=
\left[ \begin{array}{cc}
{\bf I}-{\bf B}{\bf S}^{-1}{\bf C} & {\bf B}{\bf S}^{-1} \\
{\bf S}^{-1}{\bf C} & -{\bf S}^{-1}
\end{array} \right]
\left[\begin{array}{c}
 \boldsymbol{\omega} \\ 
 {\bf c}
      \end{array}\right]\,,
\label{eq:transa}
\end{equation}
where the $(N_{\rm sq}\!-\!1)\times(N_{\rm sq}\!-\!1)$ matrix ${\bf
  S}$ is given by
\begin{equation}
{\bf  S}={\bf C}{\bf B}\,,
\label{eq:S}
\end{equation}
transforms~(\ref{eq:disc2}) into
\begin{equation}
\left[ \begin{array}{cc}
{\bf I}+{\bf R}{\bf K}^{\circ}({\bf I}-{\bf B}{\bf S}^{-1}{\bf C}) & 
{\bf R}{\bf K}^{\circ}{\bf B}{\bf S}^{-1} \\
{\bf 0}                        & {\bf I}
\end{array} \right]
\left[\begin{array}{c}
 \boldsymbol{\omega} \\ 
 {\bf c}  
      \end{array}\right]
= \left[ \begin{array}{c}
{\bf R}{\bf g} \\ 0 \end{array} \right]\,.
\label{eq:disc3}
\end{equation}

From~(\ref{eq:disc3}) it is obvious that ${\bf c}\!=\!{\bf 0}$ and we
can write~(\ref{eq:disc3}) as a single equation for
$\boldsymbol{\omega}$:
\begin{equation}
\left(
{\bf I}+{\bf R}{\bf K}^{\circ}({\bf I}-{\bf B}{\bf S}^{-1}{\bf C})
\right)\boldsymbol{\omega}={\bf R}{\bf g}\,.
\label{eq:disc4}
\end{equation}
The effective conductivity~(\ref{eq:eff2D}) can be expressed in terms
of $\boldsymbol{\omega}$ as
\begin{equation}
\sigma_{\ast}=\Im\left\{
\bar{{\bf e}}_{\rm coa}^T{\bf Z}_{\rm coa}
({\bf I}-{\bf B}{\bf S}^{-1}{\bf C})\boldsymbol{\omega}
\right\}\,.
\label{eq:eff2E}
\end{equation}

\subsection{The inverse of ${\bf S}$}
\label{sec:Sinv}

The matrix ${\bf S}$ of~(\ref{eq:S}) is sparse. In the limit
$\lambda(z)\to 0$ it approaches a standard five-point stencil for the
discrete Laplace operator. As we soon shall see, {\sc Matlab}'s sparse
banded solver, obtained using `backslash', is very efficient at
solving linear systems with ${\bf S}$ as system matrix, at least for
system sizes up to $N_{\rm sq}\!=\!1.6\cdot 10^6$. We shall use that
method in all numerical examples.

The condition number of ${\bf S}$ seems to be lower when
$\sigma(N_{\rm sq})\!=\!\sigma_2$ than when $\sigma(N_{\rm
  sq})\!=\!\sigma_1$. Therefore, in our numerical examples, we permute
the unit cell so that $\sigma(N_{\rm sq})\!=\!\sigma_2$.

\subsection{Reduction in the number of unknowns}

Some entries of $\boldsymbol{\omega}$ in~(\ref{eq:disc4}) are easy to
solve for. To see this, let $\Gamma_{\rm eq}$ be the part of
$\Gamma_0$ that lies between squares of equal conductivity.
From~(\ref{eq:lmb}) it follows that $\lambda(z)\!=\!0$ for
$z\in\Gamma_{\rm eq}$. This means that all entries of ${\bf
  K}^{\circ}$ and ${\bf R}$ whose first index corresponds to a
discretization point $z_i\in\Gamma_{\rm eq}$ are zero except for the
diagonal entries $R_{ii}$ which are one. From~(\ref{eq:disc4}) we get
the simple entrywise relation
\begin{equation}
{\omega}_i=g_i\,,\quad z_i\in\Gamma_{\rm eq}\,.
\label{eq:reduc}
\end{equation}
Furthermore, the vast majority of these elements $g_i$ are zero thanks
to $l_0(z)$, see the right hand side of~(\ref{eq:int2}).

Eq.~(\ref{eq:reduc}) can be used to reduce the number of unknowns
in~(\ref{eq:disc4}). The savings are huge when the area fraction $p$
is high or low. For simplicity, we only remove the known entries of
$\boldsymbol{\omega}$ which are zero. The reduced system assumes the
form
\begin{equation}
\left(
\downarrow\!\!\vec{{\bf I}}
+\downarrow\!\!\vec{{\bf R}}
\downarrow\!\!{\bf K}^{\circ}
(\vec{{\bf I}}-{\bf B}{\bf S}^{-1}\vec{{\bf C}})
\right)\boldsymbol{\omega}_{\rm u}=\downarrow\!\!{\bf R}{\bf g}\,.
\label{eq:disc5}
\end{equation}
Here $\boldsymbol{\omega}_u$ are the remaining entries of
$\boldsymbol{\omega}$, `downarrow' indicates that rows of a matrix are
deleted, and `rightarrow' indicates that columns are deleted. One can
see in~(\ref{eq:disc5}) that the reduction in the number of unknowns
does not induce a similar reduction in the size of ${\bf K}^{\circ}$.
No columns are deleted. Therefore, the speedup resulting
from~(\ref{eq:disc5}) is not as great as the savings in storage.

\section{Numerical examples}
\label{sec:numex}

This section investigates the complexity and the achievable accuracy
of our scheme~(\ref{eq:comp1}) for ordered arrays of squares
and~(\ref{eq:disc4}) and~(\ref{eq:disc5}) for random checkerboards. We
also compare with recent numerical results~\cite{Chen09} obtained with
the finite element solver {\sc Abacus}.

The numerical examples are performed in the {\sc Matlab} environment
(version 7.9). The GMRES iterative solver~\cite{Saad86} and a threaded
version of the fast multipole method~\cite{Gree87}, coded in C with
SIMD instructions, is used for the main linear systems. The stopping
criterion threshold is set to machine epsilon ($\epsilon_{\rm mach}$).
See Section~4.1 of~\cite{Gree87} and Section~3 of~\cite{Gree04} for
how to impose periodic boundary conditions on potential fields due to
charges in a unit cell. The examples involving $N_{\rm sq}$ up to
around $10^6$ are executed on a workstation equipped with an IntelXeon
E5430 CPU at 2.66 GHz and 32 GB of memory while all other examples are
executed on a workstation equipped with an IntelCore2 Duo E8400 CPU at
3.00 GHz and 4 GB of memory.

When estimating accuracy we rely on some exact relations available for
two-component media and compiled in Chapters~3.2 and~8.7 of
Milton~\cite{Milt02}. Let us consider $\sigma_{\ast}$ and the
effective conductivity tensor $\boldsymbol{\sigma}_{\ast}$ as
functions of $\sigma_1$ and $\sigma_2$. Then, using a duality
transform and the homogeneity of $\boldsymbol{\sigma}_{\ast}$, one can
show the following relation between an original material and that of a
material where the components have been interchanged
\begin{equation}
  \boldsymbol{\sigma}_{\ast}(\sigma_2,\sigma_1)=\sigma_1\sigma_2
  \boldsymbol{\sigma}_{\ast}(\sigma_1,\sigma_2)/
  \det(\boldsymbol{\sigma}_{\ast}(\sigma_1,\sigma_2))\,.
  \label{eq:dual}
\end{equation}
Another useful relation which holds for overall isotropic materials is
\begin{equation}
  \sigma_{\ast}(\sigma_1,\sigma_2)
  \bar{\sigma}_{\ast}(1/\bar{\sigma}_1,1/\bar{\sigma}_2)=1\,.
\label{eq:iso}
\end{equation}
An ordered checkerboard has
\begin{equation}
  \sigma_{\ast}=\sqrt{\sigma_1\sigma_2}
  \label{eq:order}   
\end{equation}
and a square array of squares at $p\!=\!0.25$ has
\begin{equation}
  \sigma_{\ast}=
  \sigma_1\sqrt{(\sigma_1+3\sigma_2)/(3\sigma_1+\sigma_2)}\,.
  \label{eq:sqsq25}
\end{equation}
We also observe, see Chapter~1.7 of~\cite{Milt02}, that the effective
conductivity of a random checkerboard at $p\!=\!0.5$ obeys
\begin{equation}
\lim_{N_{\rm sq}\to\infty}{\sigma_{\ast}}=\sqrt{\sigma_1\sigma_2}\,.
\label{eq:statlim}
\end{equation}

Note that the tensor $\boldsymbol{\sigma}_{\ast}$ has four elements
and that they all can be computed via~(\ref{eq:disc4})
(or~(\ref{eq:disc5})) and~(\ref{eq:eff2E}). For example, choosing
$e\!=\!1$ both in~(\ref{eq:disc4}), where $e$ appears in ${\bf g}$,
and in~(\ref{eq:eff2E}) makes $\sigma_{\ast}$ assume the value of
$\sigma_{\ast xx}$. Choosing $e\!=\!1$ in~(\ref{eq:disc4}) and
$e\!=\!{\rm i}$ in~(\ref{eq:eff2E}) makes $\sigma_{\ast}$ assume the
value of $\sigma_{\ast yx}$.

\begin{figure}[t]
\begin{center}
  \includegraphics[height=67mm]{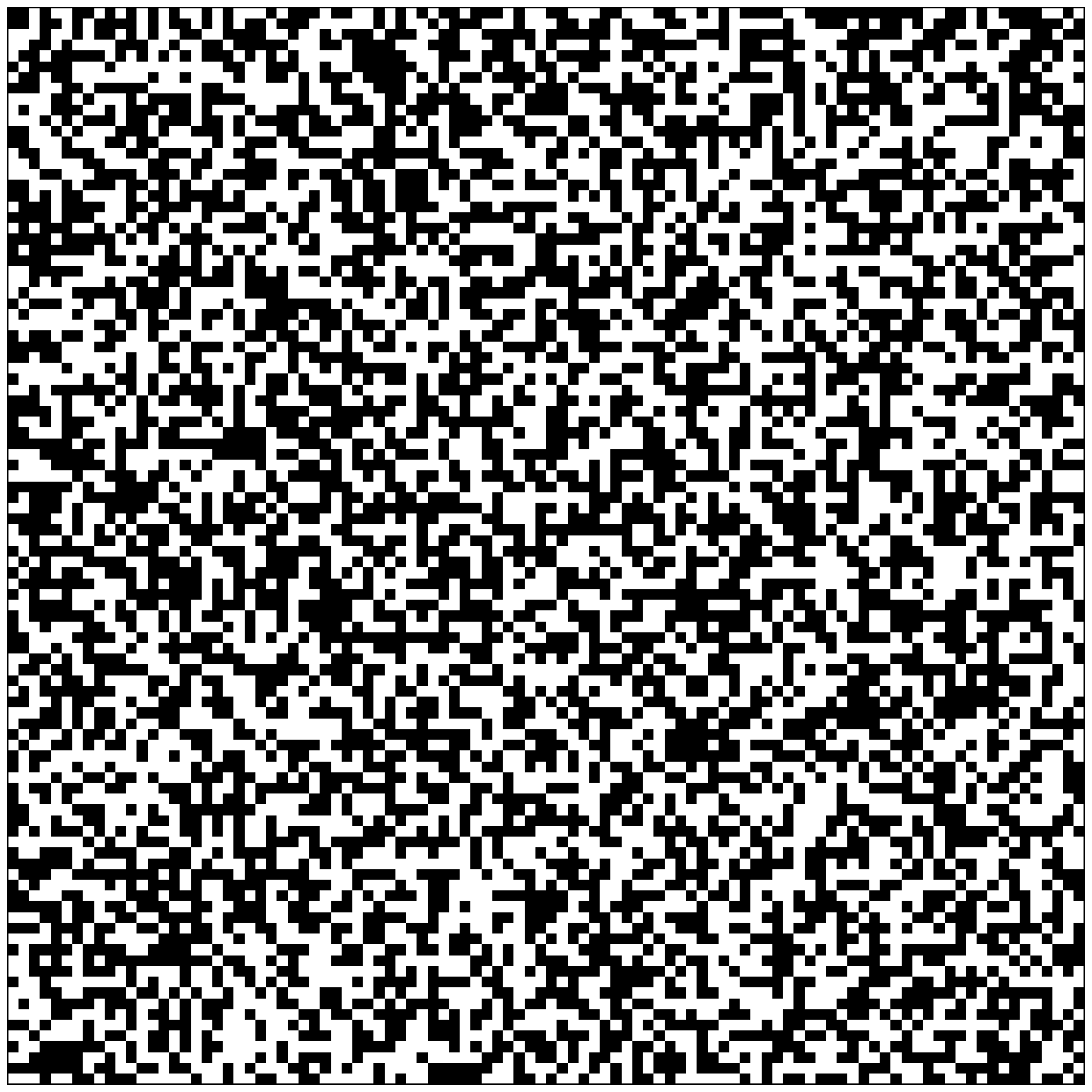}
  \includegraphics[height=67mm]{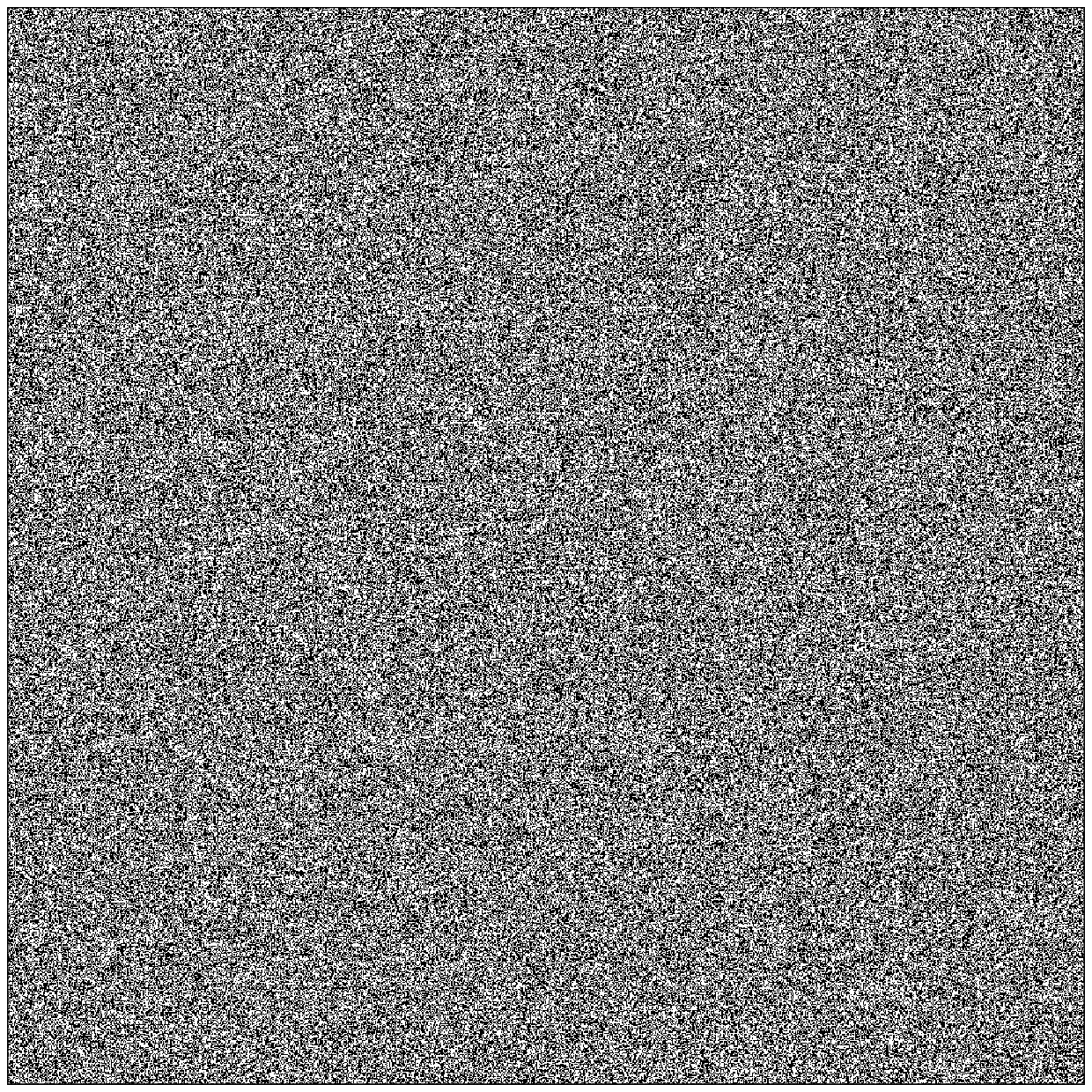}
\end{center}
\caption{\sf Unit cells of two random checkerboards used in the
  numerical examples. The area fraction of squares with conductivity
  is $\sigma_2$ is $p\!=\!0.5$. Left: $N_{\rm sq}\!=\!10^4$. Right:
  $N_{\rm sq}\!=\!10^6$. Please zoom to see the details of the
  microstructure in the right image.}
\label{fig:ch}
\end{figure}

\begin{figure}[t]
\begin{center}
  \includegraphics[height=69mm]{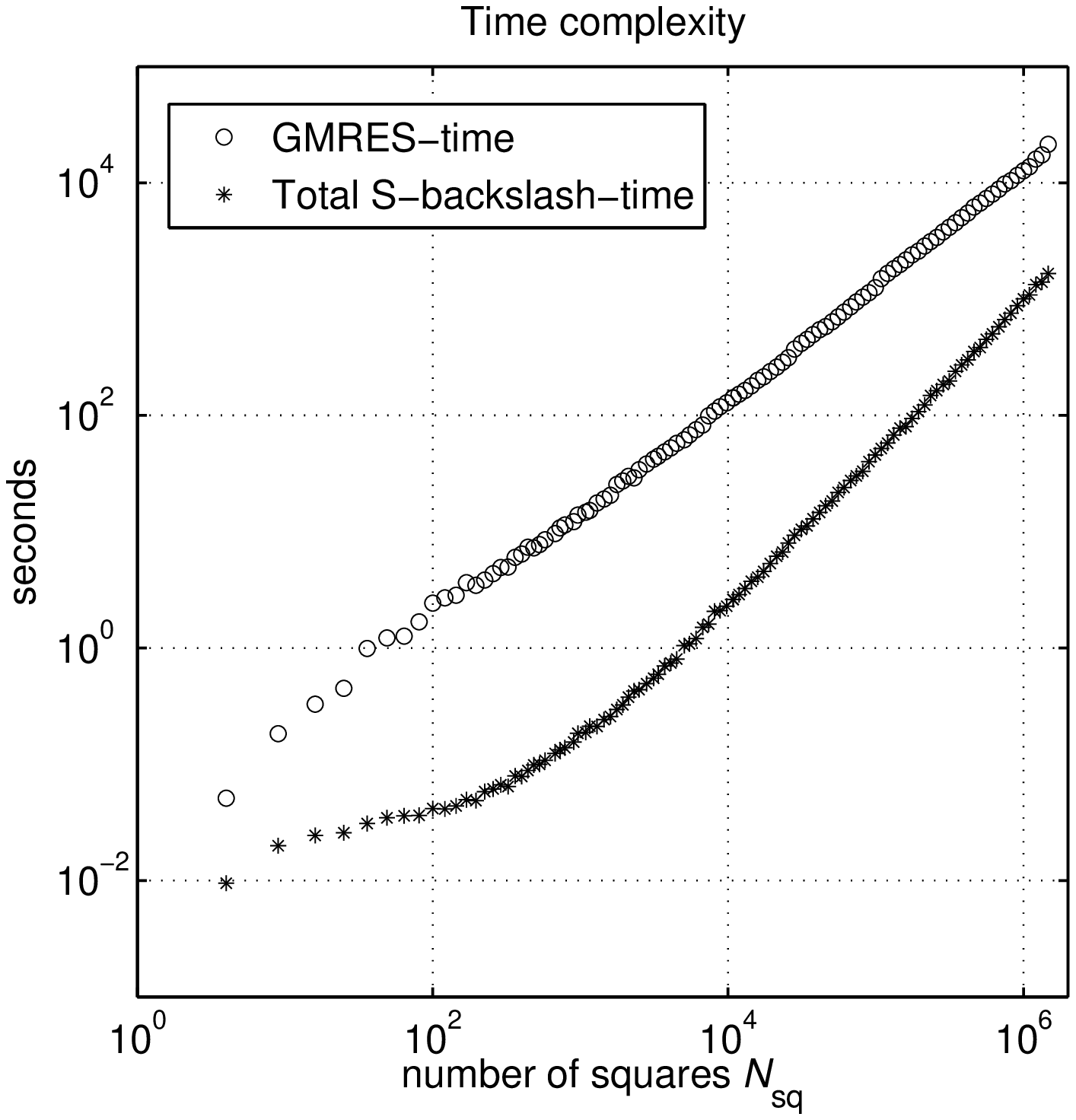}
  \includegraphics[height=69mm]{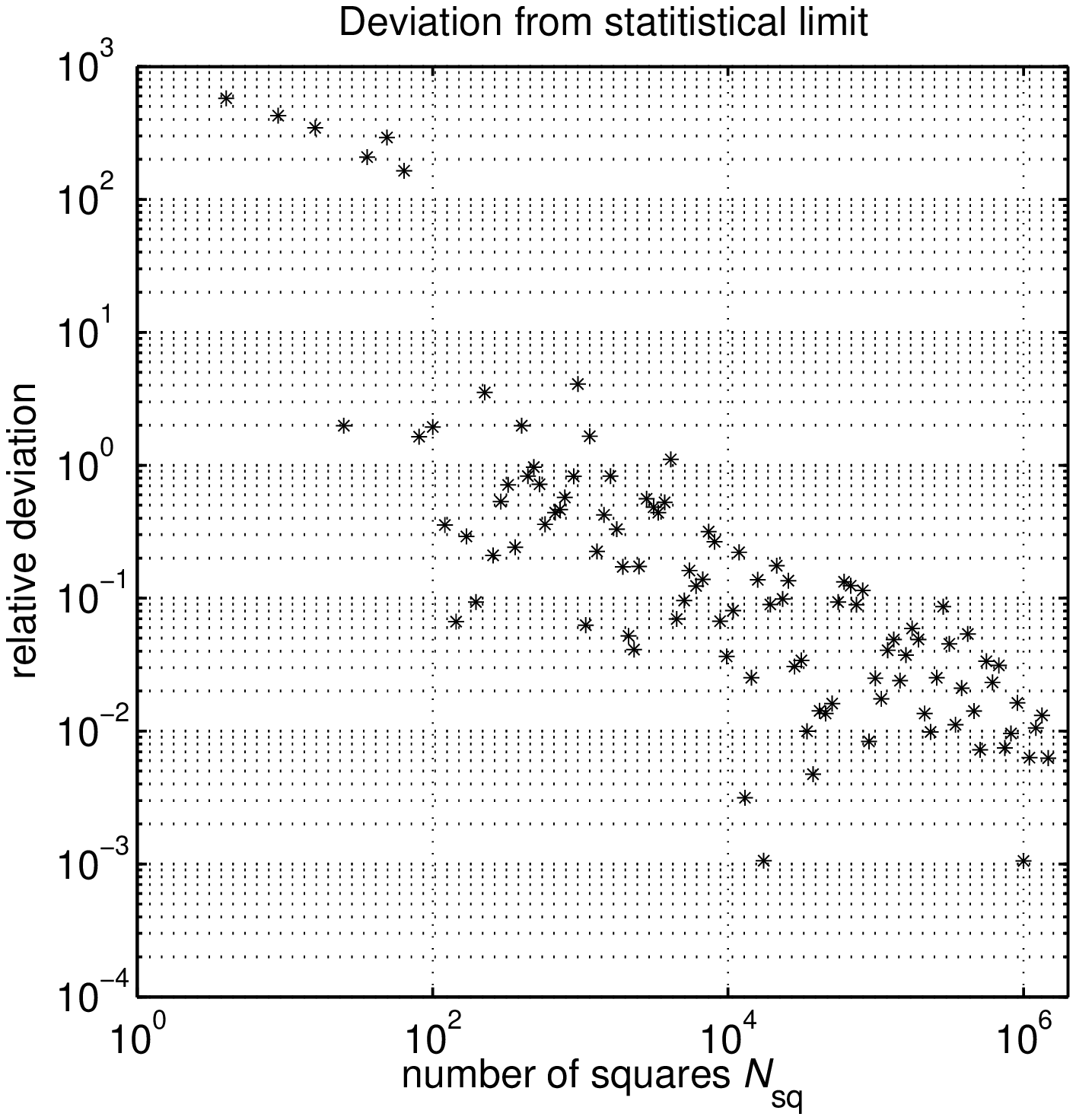}
\end{center}
\caption{\sf Solving for the effective conductivity of random
  checkerboards with $\sigma_2/\sigma_1=10^6$ and unit cells of
  increasing sizes. The area fraction of squares with conductivity
  $\sigma_2$ is $p\!=\!0.5$.}
\label{fig:time}
\end{figure}

\subsection{Timing and convergence to statistical limit}
\label{sec:statist}

A sequence of 105 random checkerboards is constructed with unit cell
sizes ranging from $N_{\rm sq}\!=\!4$ to $N_{\rm sq}\!=\!1615441$. All
unit cells have $\sigma_2/\sigma_1\!=\!10^6$ and $p\!=\!0.5$, see
Fig.~\ref{fig:ch} for two layouts. The effective conductivities
$\sigma_{\ast yy}$ of the checkerboards are computed via
(\ref{eq:disc5}) and~(\ref{eq:eff2E}).

Almost all computing time is spent in the GMRES solver. The setup time
for ${\bf S}$ of~(\ref{eq:S}) at $N_{\rm sq}\!=\!10^6$, for example,
is only about $0.4\%$ of the total computing time. The number of
iterations needed for full convergence is bounded by $18$ and the left
image of Fig.~\ref{fig:time} shows that the time spent in GMRES grows
approximately linearly with $N_{\rm sq}$, reflecting the complexity of
the fast multipole method. The total time spent applying the inverse
of ${\bf S}$, which is included in the time spent in GMRES, is also
shown separately in the left image of Fig.~\ref{fig:time}. One can see
that while this time grows faster than linearly, it is still less than
$9\%$ of the total computing time at $N_{\rm sq}\!=\!10^6$.

The right image of Fig.~\ref{fig:time} shows the actual values for the
effective conductivities $\sigma_{\ast}$ of the checkerboards,
presented in terms of their relative deviation from the statistical
limit~(\ref{eq:statlim}). At $N_{\rm sq}\!=\!1615441$, which is the
largest unit cell we can handle due to memory constraints, the
deviation is about 1\%.

\begin{figure}[t]
\begin{center}
  \includegraphics[height=68mm]{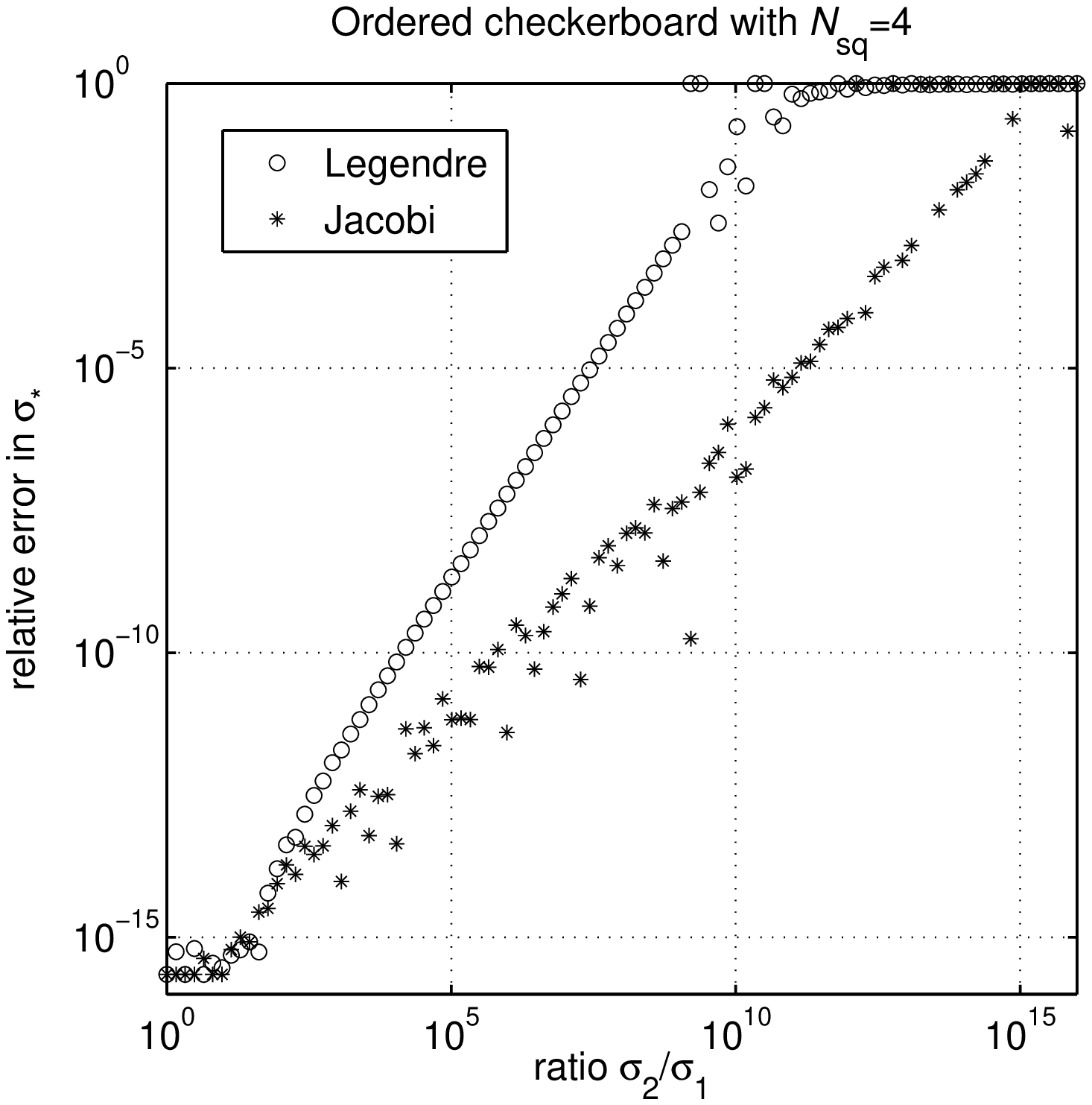}
  \includegraphics[height=68mm]{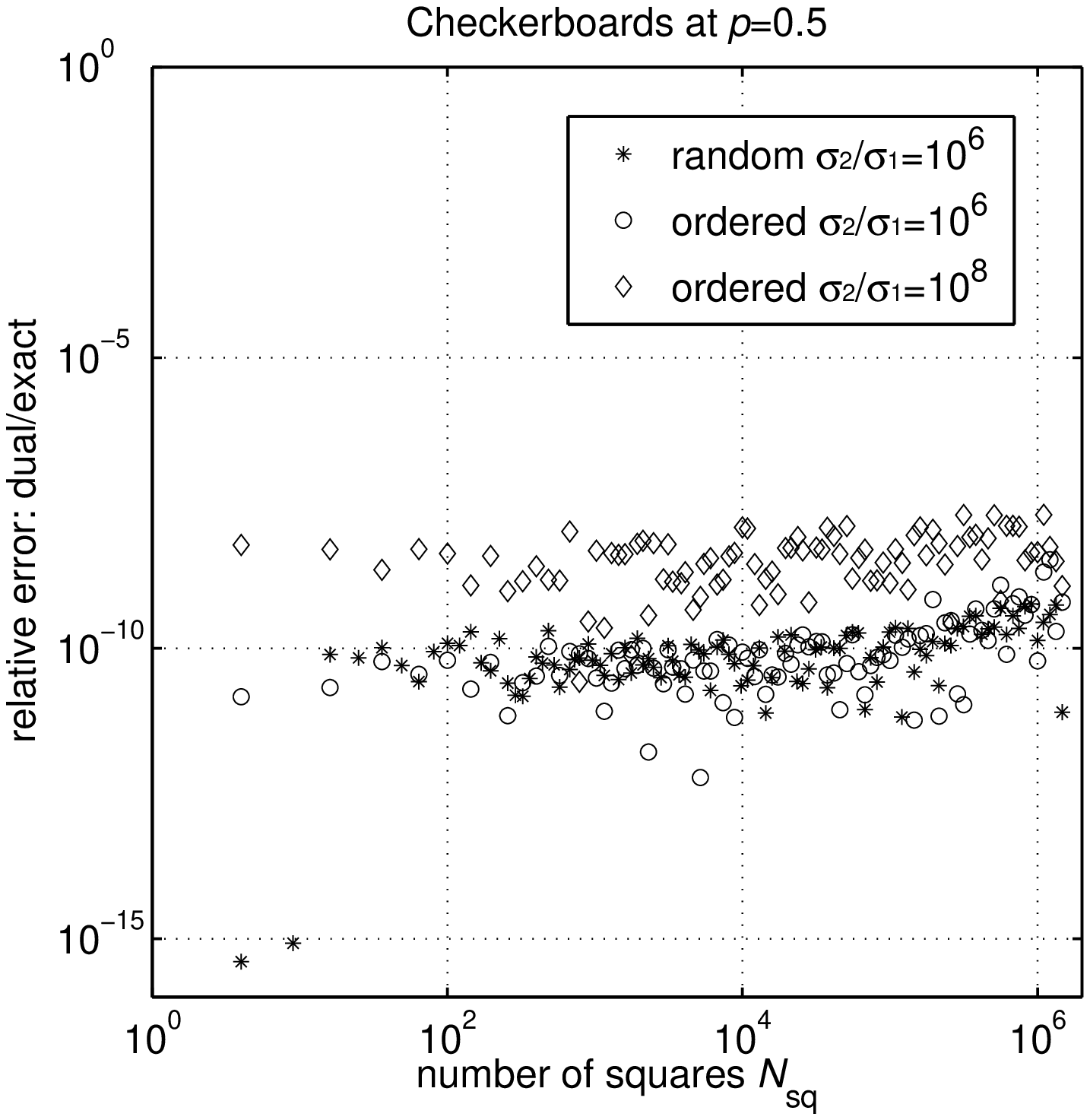}
\end{center}
\caption{\sf Achievable accuracy in the effective conductivity 
  $\sigma_{\ast}$. Left: Two strategies for placing nodes on special
  panels are compared. Errors outside of the range $[\epsilon_{\rm
    mach}, 1]$ are shown as either $\epsilon_{\rm mach}$ or $1$,
  whichever is closest. Right: error growth as a function of unit cell
  size.}
\label{fig:accu}
\end{figure}

\subsection{Achievable accuracy}
\label{sec:accu}

Fig.~\ref{fig:accu}, left image, illustrates how the placement of
quadrature nodes influences the achievable accuracy for progressively
higher contrast ratios. The unit cell is that of an ordered
checkerboard with $N_{\rm sq}\!=\!4$. The circles show that the
relative error in $\sigma_{\ast}$ grows roughly as
$(\sigma_2/\sigma_1)^{1.5}$ when Legendre nodes are used on all
panels. This was the strategy in~\cite{Hels11b}. The stars show that
the growth rate becomes linear in $\sigma_2/\sigma_1$ when Jacobi
nodes are used on special panels. This is the strategy of the present
paper, see Section~\ref{sec:disc}. Several extra digits can be
obtained at high contrast ratios.

Note that for $\sigma_2/\sigma_1>10^{16}$, accurate results are
impossible in double precision arithmetic. This is so since
$\lambda(z)$ of~(\ref{eq:int1}) and~(\ref{eq:int2}) is then
indistinguishable from unity. The integral equations become
independent of $\sigma_2$ while the reference
solution~(\ref{eq:order}) is not. The error growth rate produced by
the Jacobi nodes in Fig.~\ref{fig:accu} could therefore be thought of
as optimal.

Three sequences of checkerboards are now constructed with unit cell
sizes ranging from $N_{\rm sq}\!=\!4$ to $N_{\rm sq}\!=\!1468944$ and
with $p\!=\!0.5$. The first sequence consists of random checkerboards
with $\sigma_2/\sigma_1\!=\!10^6$. The relative errors in their
computed effective conductivities are estimated via~(\ref{eq:dual}) as
\begin{equation}
\frac{||\boldsymbol{\sigma}_{\ast}(\sigma_2,\sigma_1)-\sigma_1\sigma_2
\boldsymbol{\sigma}_{\ast}(\sigma_1,\sigma_2)/
\det(\boldsymbol{\sigma}_{\ast}(\sigma_1,\sigma_2))||_2}
{||\boldsymbol{\sigma}_{\ast}(\sigma_2,\sigma_1)||_2}\,.
\label{eq:dualerr}
\end{equation}
The second sequence consists of ordered checkerboards with
$\sigma_2/\sigma_1\!=\!10^6$ and~(\ref{eq:order}) is used as reference
solution. The third sequence is the same as the second sequence, but
the contrast ratio is increased to $\sigma_2/\sigma_1\!=\!10^8$.

Fig.~\ref{fig:accu}, right image, shows the results and it has several
interesting features. For example, one can see that:
\begin{itemize} 
\item the error in $\sigma_{\ast}$ seems to be independent of the unit
  cell size. This is so because the total error is dominated by the
  error caused by corner self-interaction, computed in local
  coordinates. The error from long-range interaction is comparatively
  small except for $N_{\rm sq}> 5\cdot 10^5$.
\item the error estimate for $\boldsymbol{\sigma}_{\ast}$ of random
  checkerboards~(\ref{eq:dualerr}), based on duality, agrees well with
  the error estimate for $\sigma_{\ast}$ of ordered checkerboards,
  based on an exact answer~(\ref{eq:order}).
\end{itemize}

\begin{figure}[t]
\begin{center}
  \includegraphics[height=56mm]{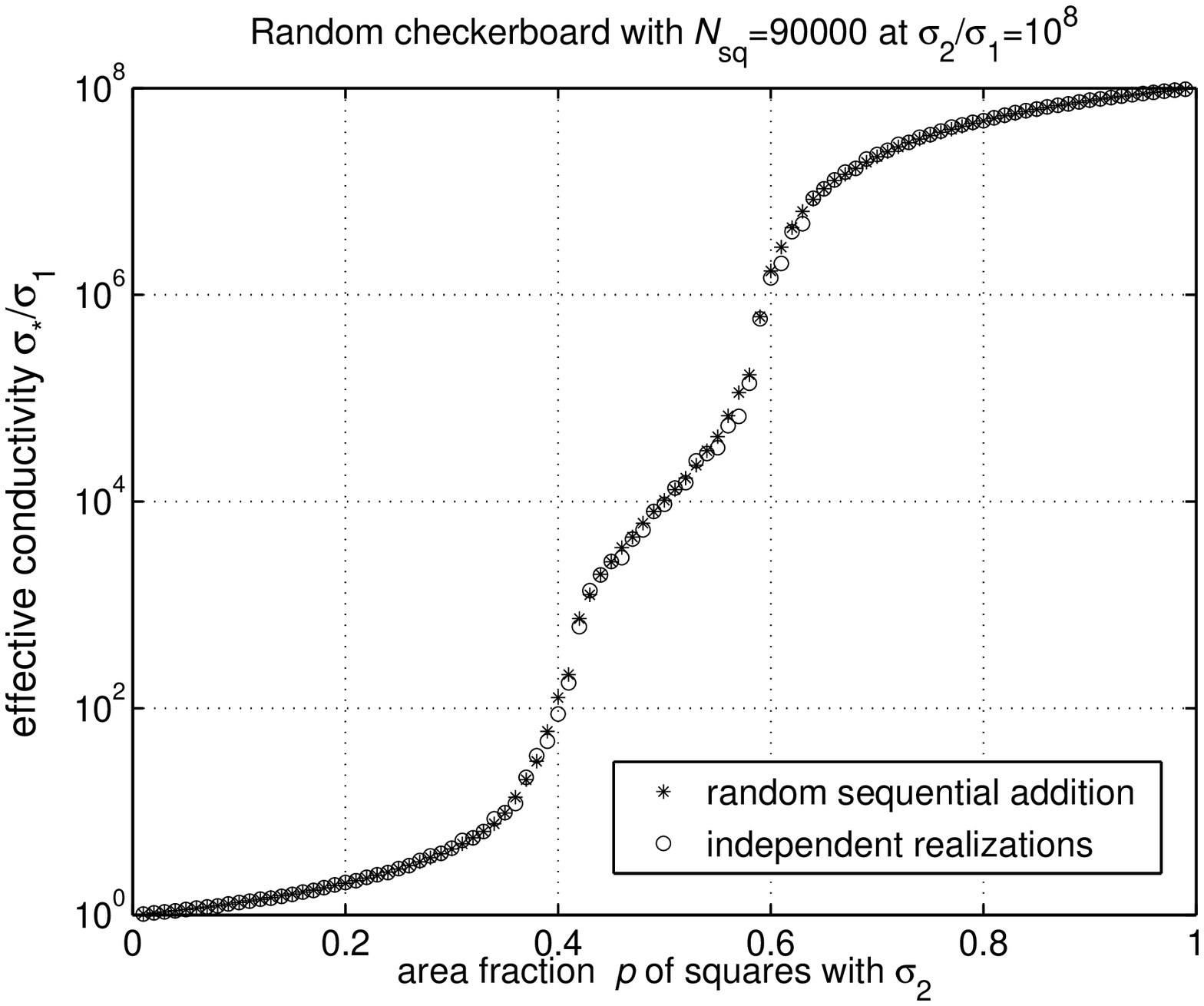}
  \includegraphics[height=56mm]{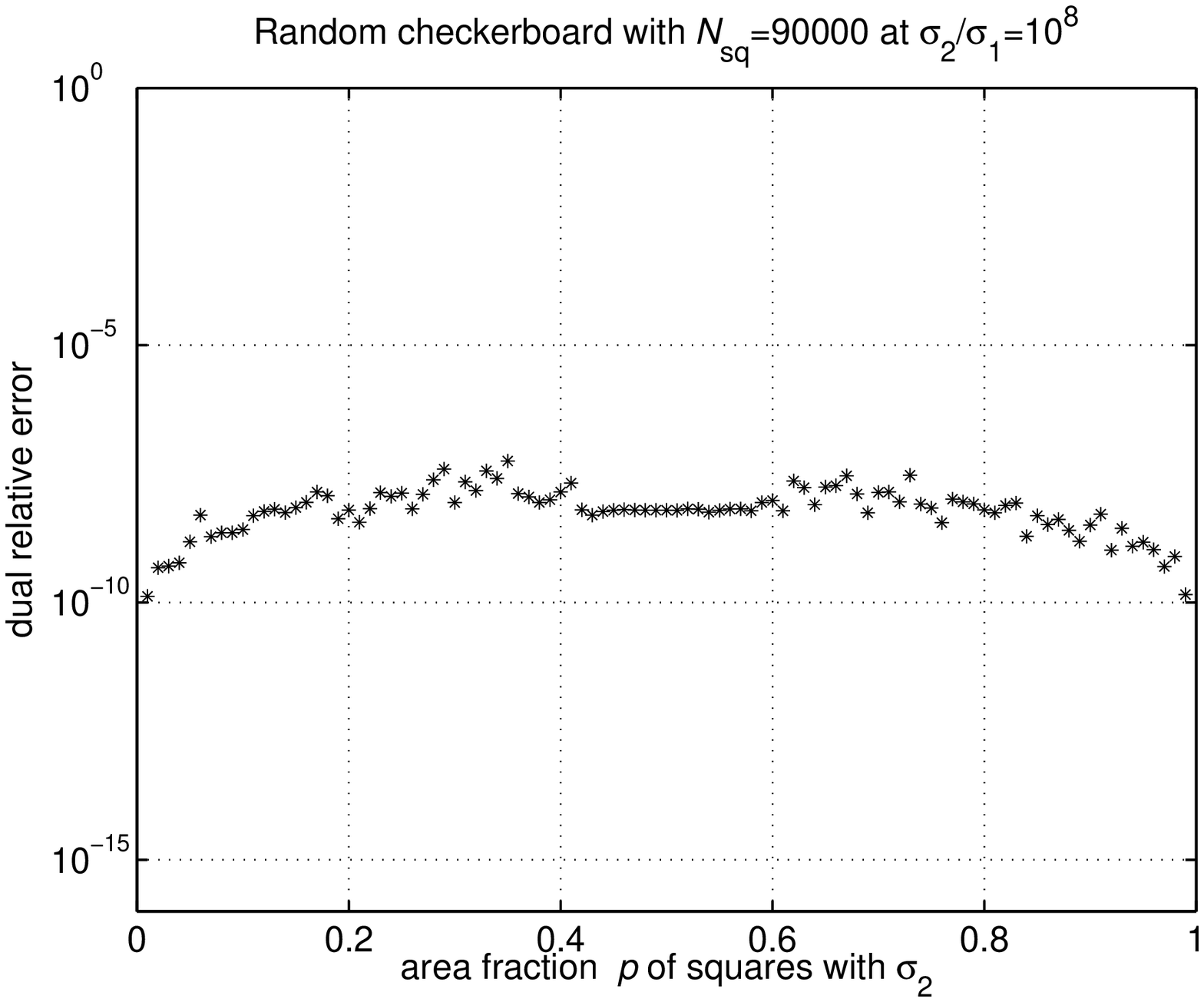}
\end{center}
\caption{\sf Random checkerboards with $N_{\rm
    sq}=90000$ squares in the unit cell and $\sigma_2/\sigma_1=10^8$.
  The area fraction of squares with conductivity $\sigma_2$ varies
  from $p=0$ to $p=1$.}
\label{fig:schuh}
\end{figure}

\subsection{Continuum percolation}

In theoretical materials science it is of interest to study the
effective conductivities of continuum two-component random composites
as $p$ varies. Fig.~\ref{fig:schuh} shows such a study for a unit cell
with $N_{\rm sq}\!=\!90000$ and $\sigma_2/\sigma_1\!=\!10^8$ along
with the error estimate~(\ref{eq:dualerr}). Two sequences of
realizations are shown -- one based on sequential random addition and
one where all realizations are independent. It is obvious, from the
jagged shape of the curve in the left image and also from the results
in Section~\ref{sec:statist}, that we are far from the statistical
limit. Percolation thresholds are visible at $p\approx 0.41$ and at
$p\approx 0.59$. These numbers are consistent with classic results on
site percolation for a square lattice~\cite{Lee08}. The overall
behavior of $\sigma_{\ast}$ as a function of $p$ in
Fig.~\ref{fig:schuh} is in agreement with the discussion on p.~207 in
Milton~\cite{Milt02} and also with results obtained with a discrete
network model~\cite{Mart03} but it stands in contrast to results
obtained with the finite element method in Fig.~2(a) of~\cite{Chen09}.
There only one percolation threshold is observed.

\begin{figure}[t]
\begin{center}
  \includegraphics[height=55mm]{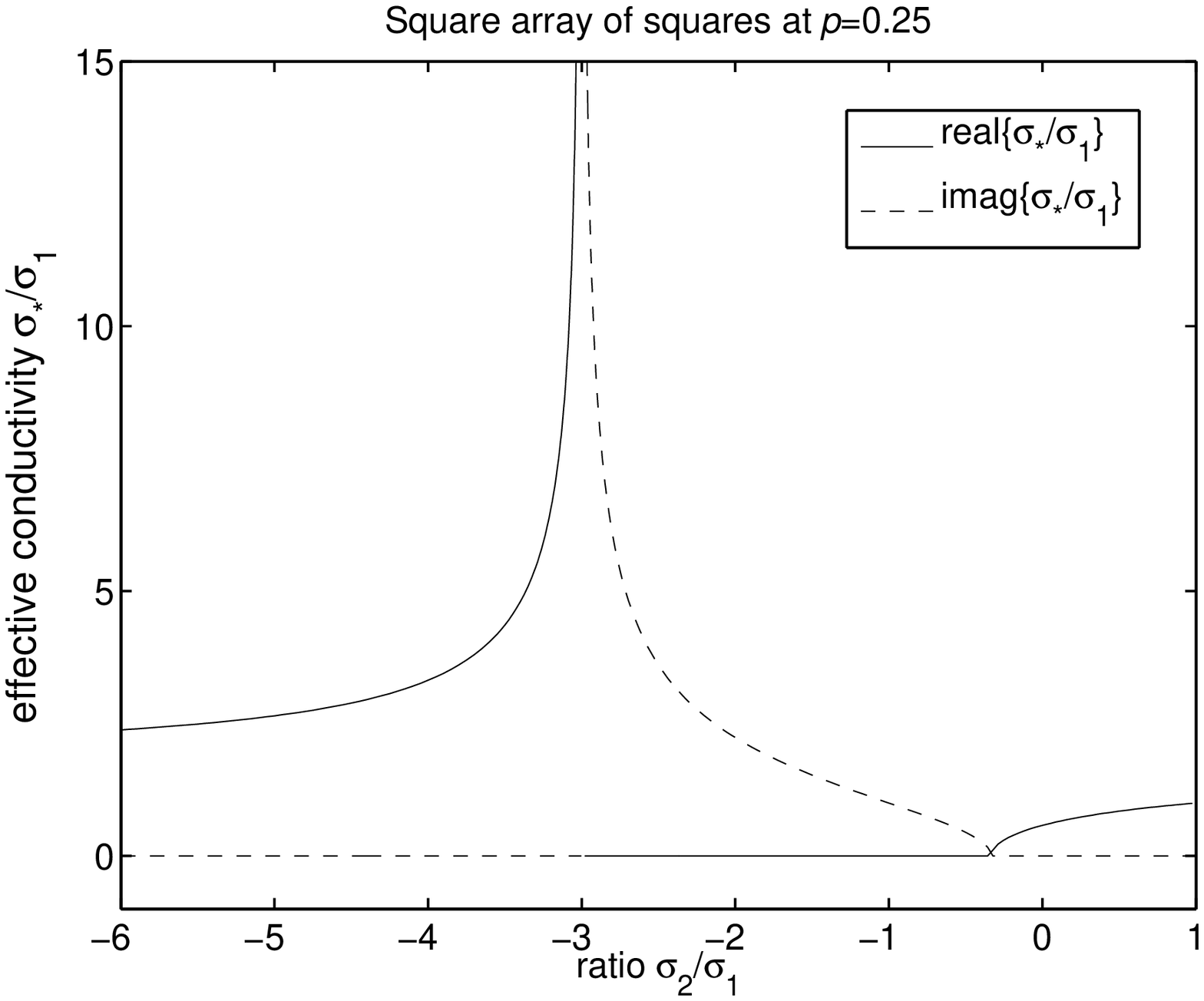}
  \includegraphics[height=55mm]{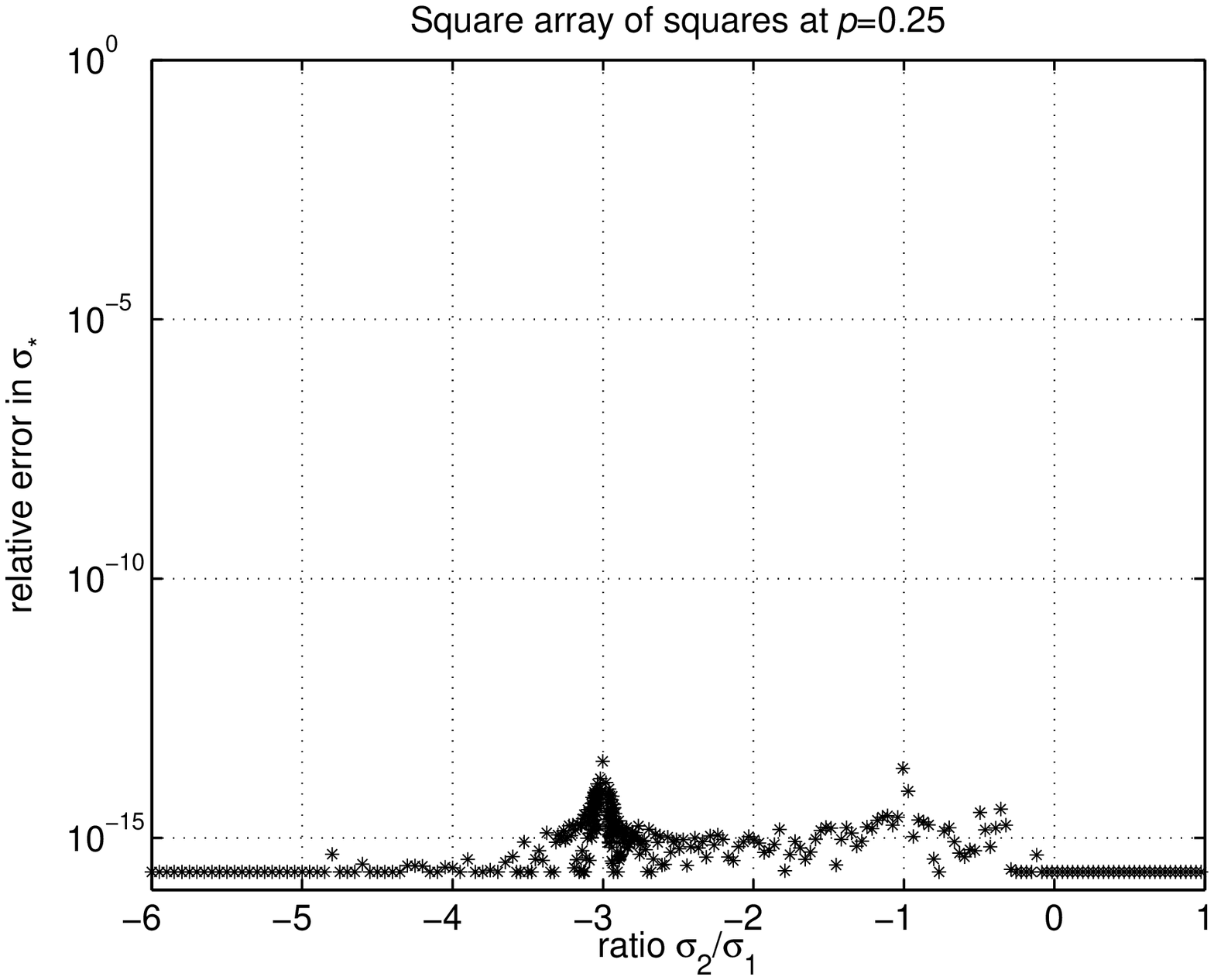}
\end{center}
\caption{\sf Left: the effective conductivity $\sigma_{\ast}/\sigma_1$
  of a square array of squares at $p=0.25$. The curves are supported
  by 349 adaptively spaced data points. Right: the relative error
  with~(\ref{eq:sqsq25}) as reference solution.}
\label{fig:sqsq25}
\end{figure}

\subsection{The square array of squares}
\label{sec:sqsq}

Fig.~\ref{fig:sqsq25} shows computed values of
$\sigma_{\ast}/\sigma_1$ for the square array of squares at $p=0.25$
for negative ratios $\sigma_2/\sigma_1$. The relative error,
with~(\ref{eq:sqsq25}) as reference solution, is shown in the right
image. The error is close to $\epsilon_{\rm mach}$ except for in a
neighborhood of three points where it is higher: the `pole' or
`resonance' at $\sigma_2/\sigma_1=-3$, the `essential singularity' at
$\sigma_2/\sigma_1=-1$, and the `zero' at $\sigma_2/\sigma_1=-1/3$.
See~\cite{Perr10} for an explanation of the significance and physical
meaning of these terms. Note that at $\sigma_2/\sigma_1=-1$ we have
$\lambda=\pm\infty$ and that~(\ref{eq:int1}) then becomes a first kind
equation. Compare also Fig.~2 of~\cite{Perr10}, which is similar to
the left image of our Fig.~\ref{fig:sqsq25}, but where some problems
are encountered along the branch cut $\sigma_2/\sigma_1\in[-3,-1/3]$.

\begin{figure}[t]
\begin{center}
  \includegraphics[height=56mm]{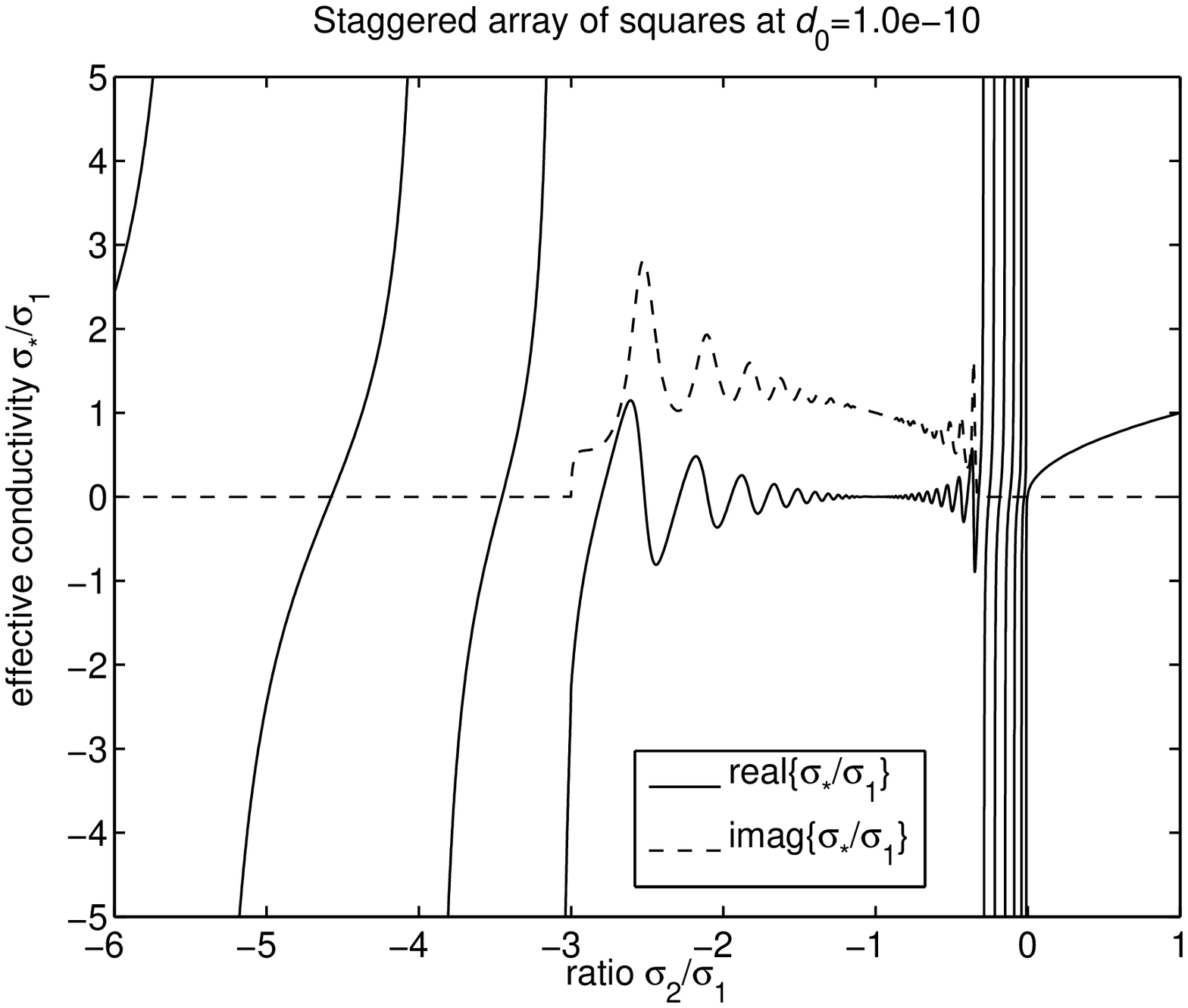}
  \includegraphics[height=56mm]{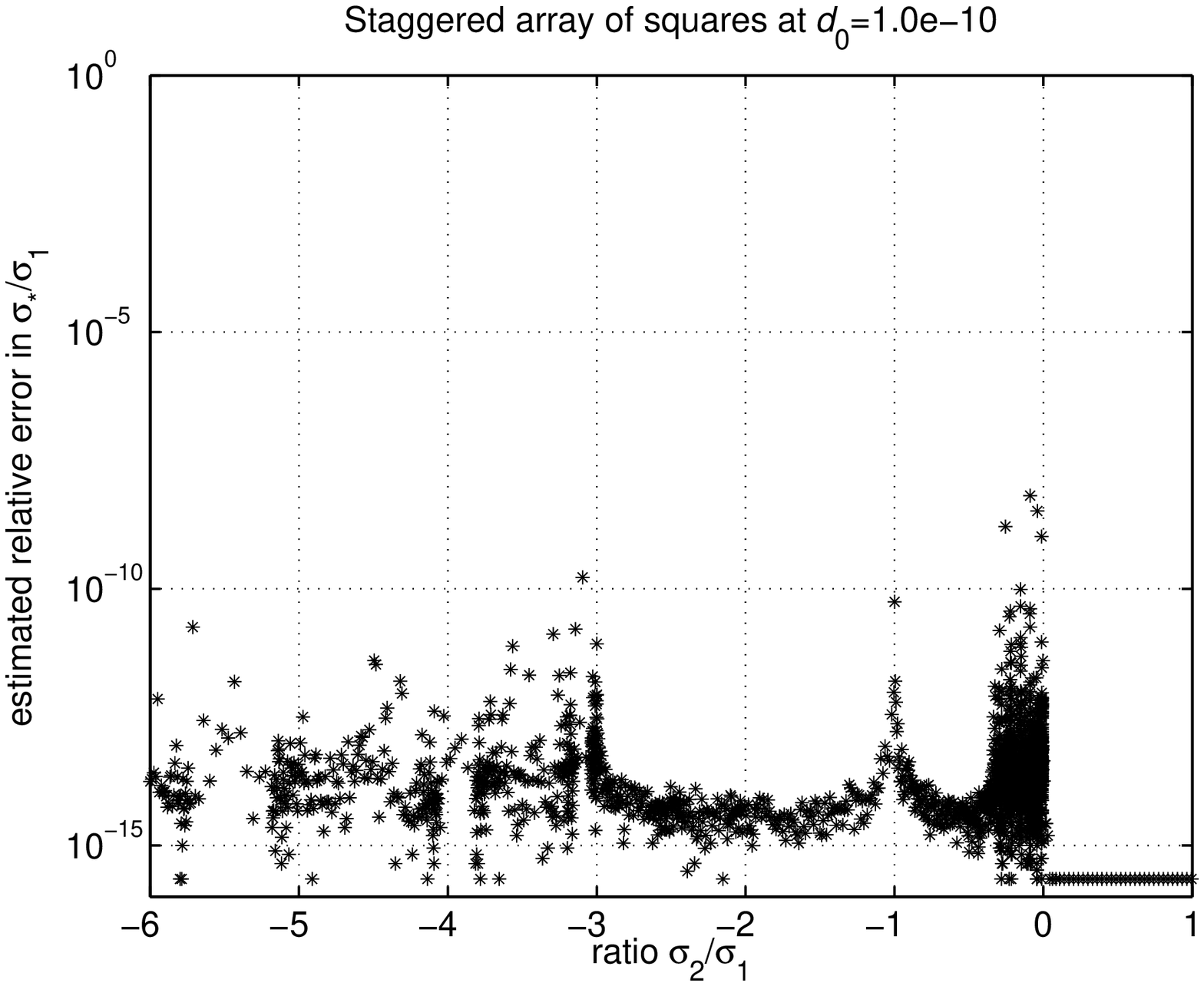}
\end{center}
\caption{\sf The effective conductivity $\sigma_{\ast}/\sigma_1$ of 
  a staggered array of squares at $d_0\!=\!10^{-10}$. The curves are
  supported by 2006 data points. Right: an error estimate based
  on~(\ref{eq:iso}).}
\label{fig:stagex}
\end{figure}

\subsection{The staggered array of squares}

Staggered arrays of squares at area fractions close to $p\!=\!0.5$
exhibit rich resonant spectra on the negative real axis and pose
greater challenges to numerics than the example of
Section~\ref{sec:sqsq}. More data points are required to resolve
$\sigma_{\ast}$. For modeling purposes it is convenient to describe
staggered arrays in terms of a parameter $d_0$, related to the area
fraction $p$ and to the vertex separation distance $d$, see
Figs.~\ref{fig:F1} and~\ref{fig:stagmesh}, as
\begin{equation}
p=\frac{2}{(d_0+2)^2} \qquad {\rm and} \qquad d=\frac{d_0}{\sqrt{2}(d_0+2)}\,.
\end{equation}

The left image of Fig.~\ref{fig:stagex} for $d_0\!=\!10^{-10}$ shows
an oscillatory behavior of $\sigma_{\ast}/\sigma_1$ for
$\sigma_2/\sigma_1\in[-3,-1/3]$ and a number of resonances on the
negative real axis outside of this interval. The right image shows
that the relative error in these computations, estimated via how
well~(\ref{eq:iso}) is met, is typically on the order of
$10^2\epsilon_{\rm mach}$. Close to the eigenvalues
of~(\ref{eq:int1}), some of which correspond to poles of
$\sigma_{\ast}$, the error is of course larger. The largest relative
error encountered in this example is estimated to $10^{-8}$.

\section{Conclusions}

The homogenization of composite materials with large random unit cells
of squares at extreme material property ratios is a canonical problem
in the theory of composite materials. It has fascinated researchers
for decades~\cite{Milt02}. The domains look simple, yet they are
intriguing. There are analytical results available for special cases,
yet numerical solvers run into trouble. Only a few years ago,
numerical solutions to the type of homogenization problems presented
in this paper would be considered far out of reach.

The present work epitomizes and stretches a recent line of
research~\cite{Hels09b,Hels11a,Hels11b,Hels08,Hels09a} to a new high.
We first show how to treat simple unit cells with (almost) optimal
accuracy using a short-range preconditioner. We then show that larger
unit cells pose no extra problems when a new long-range preconditioner
is added. Our algorithm has (almost) linear complexity in both
execution time and storage requirement. Problems involving unit cells
with a million of squares can be solved to very high precision in a
few hours. Homogenization on checkerboard-like domains have become a
simple task.

How useful is our new scheme? The coupling of checkerboard problems to
real-world physics is elusive. One may question the relevance of the
small length-scales needed for the resolution of various singular
fields. Nevertheless, a recent surge in physicists' interest in
metamaterials has given new momentum to the study of these
issues~\cite{Perr10}. The difficulties arising in random checkerboard
problems may, further, be representative of the sort of troubles that
arise in several real-world problems. Since integral equation methods
are widely applicable, it is therefore likely that our scheme and
generalizations thereof have many immediate applications. Further work
based on the present scheme and directed towards metamaterial
applications is in progress~\cite{Hels11c}.

\end{document}